\documentclass[%
 amsmath, amssymb, aps,
 reprint,
 superscriptaddress,
 preprintnumbers,
 prd,
nofootinbib,
]{revtex4-2}

\usepackage{orcidlink}
\usepackage[nowatermark]{fixmetodonotes}
\usepackage[title]{appendix}
\hypersetup{colorlinks = true, allcolors = ., urlcolor = blue}

\usepackage{graphicx, tabularx}% Include figure files
\usepackage{float}
\usepackage{dcolumn}% Align table columns on decimal point
\usepackage{bm}% bold math

\usepackage[normalem]{ulem}
\usepackage{amsmath}
\usepackage{mhchem}
\newcommand{\stkout}[1]{\ifmmode\text{\sout{\ensuremath{#1}}}\else\sout{#1}\fi}

\begin{document}

\title{Implementation of a relativistic distorted wave impulse approximation model into the NEUT event generator}

\author{J.~McKean\,  \orcidlink{0009-0005-6100-6195}\,}
 \email{j.mckean21@imperial.ac.uk}
 \affiliation{Imperial College London, Department of Physics, London SW7 2BZ, United Kingdom}

\author{R. Gonz\'alez-Jim\'enez\, \orcidlink{0000-0002-0492-0619}\,}
 \email{raugj@us.es}
 \affiliation{Departamento de Física Atómica, Molecular y Nuclear, Universidad de Sevilla, 41080 Sevilla, Spain}

\author{M. Kabirnezhad, 
\orcidlink{0000-0002-3217-2259}\,}
 \email{m.kabirnezhad@imperial.ac.uk}
 \affiliation{Imperial College London, Department of Physics, London SW7 2BZ, United Kingdom}

\author{J. M. Ud\'ias\, \orcidlink{0000-0003-3714-764X}\,}
 \email{jmudiasm@ucm.es}
 \affiliation{Grupo de F\'isica Nuclear, Departamento de Estructura de la Materia, F\'isica T\'ermica y Electr\'onica, Facultad de Ciencias F\'isicas, Universidad Complutense de Madrid and IPARCOS, CEI Moncloa, Madrid 28040, Spain}

\author{Y. Uchida
\orcidlink{0000-0002-8677-9945}\,}
 \email{yoshi.uchida@imperial.ac.uk}
 \affiliation{Imperial College London, Department of Physics, London SW7 2BZ, United Kingdom}

\begin{abstract}
\noindent We describe the implementation of a model for charged-current quasi-elastic (CCQE) neutrino-nucleus scattering in the NEUT Monte Carlo event generator. 
This model employs relativistic momentum distributions obtained from mean field theory and relativistic distorted waves to describe the initial and final nucleon states. 
Final state interactions, both elastic and inelastic, are modelled by combining distorted waves with NEUT’s intranuclear cascade, offering a more accurate representation of the interactions experienced by scattered nucleons. The model and its implementation in NEUT are described in detail and benchmarked against $\nu_{\mu}$-$\ce{^{12}C}$ scattering cross-section measurements from T2K and MINER$\nu$A, as well as $\nu_{\mu}$-$\ce{^{40}Ar}$ measurements from MicroBooNE. Results, including transverse kinematic imbalance variables and scattered nucleon kinematics, show improved $\chi^2$ values compared to other CCQE models in NEUT. Furthermore, the model consistently predicts lower cross sections in CCQE-dominated regions, indicating potential for further refinement, such as incorporating two-body currents or the use of more advanced nucleon axial form factors consistent with lattice QCD calculations. 

\end{abstract}

\maketitle

\section{Introduction}
\label{sec:intro}
\noindent Neutrino-nucleus interaction modelling, particularly the treatment of nuclear effects, represents a significant contribution to the overall systematic uncertainty in neutrino oscillation parameter measurements for many long-baseline neutrino oscillation experiments~\cite{T2K-OA2023-long-paper}. In current accelerator-based neutrino oscillation experiments, such as T2K~\cite{the-t2k-exp}, statistical uncertainties dominate the errors. However, in future experiments such as Hyper-Kamiokande (HK)~\cite{HK-TDR}, DUNE~\cite{DUNE-TDR}, and the upgraded T2K era~\cite{t2k-upgrade-TDR}, systematic uncertainties stemming from neutrino interaction models are anticipated to surpass statistical uncertainties as the leading source of error. This transition highlights the critical need to reduce systematic uncertainties, which is essential for achieving precise measurements of cross sections and neutrino oscillation parameters.\\
 
\noindent For T2K and MicroBooNE~\cite{microboone-exp}, the dominant interaction channel is the charged current quasi-elastic (CCQE) interaction due to the energy at which the neutrino flux peaks. This paper focuses on CCQE interactions, where accurately modelling the kinematics of scattered nucleons is essential for a reliable description of the interaction. Key nuclear effects impacting CCQE modelling include Fermi motion, describing the intrinsic motion of nucleons within the nucleus; final state interactions (FSI), involving the re-interaction of outgoing hadrons with the residual nucleus; and multi-nucleon interactions, where neutrinos interact with correlated nucleon pairs bound within the nuclear medium.\\

\noindent A comprehensive description of FSI between hadrons and nuclei across the vast phase space explored in neutrino experiments represents an unprecedented challenge for nuclear theory. Currently, there is no microscopic theory capable of addressing this intricate coupled-channel problem. Neutrino event generators, such as NEUT~\cite{NEUT-generator}, GENIE~\cite{genieneutrinomontecarlo:2015}, ACHILLES~\cite{achilles:isaacson_2023} and NuWro~\cite{Nuwro:Golan:2012}, use the classical or semiclassical intra-nuclear cascade approach to simulate FSI. GiBUU~\cite{Buss_2012} is a different event generator based on quantum-kinetic transport theory for the propagation of hadrons in nuclear matter, which are under the influence of a real potential.\\

\noindent In Monte Carlo neutrino event generators, the cascade interaction for the CCQE reaction channel is modelled as follows: i) A nucleon from an accepted event with a given four-momentum is generated by a model of the primary CCQE vertex; ii) this nucleon is placed within the nucleus; iii) the nucleon propagates through the nucleus in temporal or spatial steps; iv) this nucleon can interact with other nucleons which can be knocked out; v) hadrons other than nucleons can also be produced in the collisions and they are also propagated in steps. The result of this process is a complete prediction of the multiplicity of hadrons in the final state, including the kinematics of each particle.
Cascade models are unitary in the sense that they do not affect the inclusive cross section, but only the composition of the hadronic final state. In other words, after summation and integration over all hadronic final states, the inclusive cross section is recovered from the primary vertex model. Therefore, it is important to start with a model for the primary vertex that gives good results for the inclusive cross section.\\

\noindent Many of the theoretical models aiming at modelling QE interactions focus on the inclusive cross section. This is the case for superscaling-based approaches~\cite{Gonzalez-Jimenez14,Amaro24}, {\it ab initio} models~\cite{Nikolakopoulos24a, Barbieri19,Pastore20,Sobczyk21}, local Fermi gas approaches~\cite{Martini10,Nieves11} and the most recent work based on neural networks, which are trained on inclusive electron scattering data~\cite{AlHammal23, Kowal24,Sobczyk24}.
The widely used local Fermi gas~\cite{Bourguille21} can provide information about the scattered nucleon, but its description of the initial and final states, based on the use of plane waves, is oversimplified.
The factorised spectral function (SF) approach~\cite{Ankowski15, Antonov11, Ivanov14, Franco-Patino20} starts with the plane-wave impulse approximation but includes a sophisticated description of the initial state, incorporating nuclear effects beyond the mean field. However, the final nucleon is described by a plane wave, i.e. the influence of the mean field, which is present in the initial state, is neglected in the final state. This lack of consistency between the description of initial and final states makes this approach unable to describe the inclusive or exclusive cross sections satisfactorily. In Ref.~\cite{Ankowski15} a ``folding function'' is used to approximately incorporate the effect of the mean field in the final state. This approach is only for inclusive cross sections and does not solve the Pauli blocking problem, which is usually imposed using different approaches.
\\

\noindent The model employed in this work is relativistic, unfactorised and provides information about the scattered nucleon.
It incorporates relativistic mean field (RMF) theory to describe the bound state~\cite{Walecka-model} and uses the relativistic distorted wave impulse approximation (RDWIA) for the final state. Various models based on this approach have been previously developed by different groups to describe inclusive and exclusive electron-nucleus and neutrino-nucleus scattering processes~\cite{Udias93, Udias01, Ivanov16, Butkevich12, Gil21, Gil22, Boffi93, Giusti11, Meucci03, Gonzalez-Jimenez19, Franco-Patino22, Franco-Patino24, Fissum04}. \\

\noindent Comparisons with electron scattering data demonstrate that this relatively simple mean-field approach consistently accounts for the bulk of nuclear effects, including shell structure, binding energies, Fermi motion, the Pauli exclusion principle, and elastic FSI, all in a quantum mechanical framework. However, the limitations of this approach are also evident; for instance, the absence of correlations necessitates the use of spectroscopic factors lower than one to reproduce experimental data accurately. To address some of these beyond-mean-field effects, the model used here incorporates a phenomenological spectral function inspired by~\cite{Benhar94,Benhar05}, while avoiding reliance on plane wave approximations. Further details of the model can be found in Refs.~\cite{Gonzalez-Jimenez22, Franco-Patino22, Franco-Patino24} and in Sect.~\ref{sec:theory}.\\

\noindent The approach employed here, consisting of combining a RDWIA model with a cascade, has been studied in Refs.~\cite{Nikolakopoulos22} and \cite{Nikolakopoulos24b}. In Ref.~\cite{Nikolakopoulos22}, protons were generated with a CCQE model based on a model that accounts for its distortion using a relativistic real optical potential. These protons were propagated through the nucleus using the NEUT cascade. The goal was to benchmark the NEUT cascade by comparing the transparency results from NEUT with the results of a well-tested microscopic model that uses a complex optical potential extracted from elastic proton-nucleus scattering data~\cite{Cooper-potentials}, so the transparency arises from the imaginary part of the potential. 
A similar exercise was done in Ref.~\cite{Nikolakopoulos24b} but in this case the cascade models from the generators NuWro, INCL~\cite{INCL13}, GENIE, ACHILLES and also NEUT were employed. \\

\noindent We emphasise that in the studies presented in \cite{Nikolakopoulos22} and \cite{Nikolakopoulos24b} the distorted-wave models were not implemented in NEUT nor any of the other generators and that only the cascade part of the generators was used, as a separate tool, to propagate the nucleon. In the present work, the distorted-wave models have been implemented fully, and for the first time, in a neutrino event generator. \\

\noindent This paper describes the first implementation of this model into the NEUT event generator framework for use on carbon, oxygen and argon. In Section~\ref{sec:theory}, the theoretical model and kinematics are introduced and the implementation of the model is described in Section~\ref{sec:implementation}. Comparisons to semi-inclusive data on carbon is shown in Section~\ref{sec:T2K}~and~\ref{sec:Minerva}, and on argon in Section~\ref{sec:uBooNE}.

\section{Theoretical framework}
\label{sec:theory}
\noindent The neutrino-nucleus interaction is a complex many-body problem which may be described in the first-order Born approximation, meaning there is only a single boson exchanged in the interaction~\cite{born-approx-max-born, landau-born-approx}. The impulse approximation (IA) is also employed such that the neutrino is considered to interact with a single bound nucleon which is then ejected~\cite{impulse-approx}. The Born approximation also allows the factorisation of the leptonic and hadronic tensors, essentially separating the lepton kinematics from the hadron kinematics. The treatment of the scattered nucleon can either include or exclude interactions with the nuclear potential as it exits the nuclear medium. Within this framework, it is possible to model elastic FSI, where the scattered nucleon interacts within the nucleus without producing any new particles. Such processes are not currently accounted for in existing Monte Carlo event generators.\\

\begin{figure}[H]
    \centering
    \includegraphics[keepaspectratio, width=0.4\textwidth]{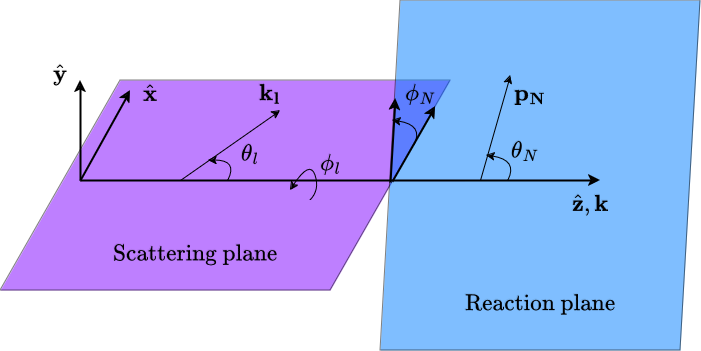}
    \caption{Schematic showing the coordinates and kinematic vectors used in this model. The scattering plane and reaction plane are explicitly separated and are chosen to contain the scattered lepton and nucleon respectively. The beam direction is always considered to be parallel to the z-axis.}
    \label{fig:ccqe-kinematics}
\end{figure}

\noindent The definition for the kinematics and the interaction vectors are shown in Fig.~\ref{fig:ccqe-kinematics}. The incident neutrino is assumed to be parallel to the z-axis, $\hat{\mathbf{z}}$, with four momentum $K^{\mu} = (E_{\nu}, \mathbf{k})$. The interaction on a single nucleon with four momentum $P^{\mu}_{A} = (M_{A}, \mathbf{p})$ produces a lepton and hadron in the final state. The leptonic part of the interaction is contained with the scattering plane, within which the scattered lepton has four momentum $K^{\mu}_{l} = (E_{l}, \mathbf{k_{l}})$ and the polar angle $\theta_{l}$ is defined as the angle between the vector $\mathbf{k_{l}}$ and $\hat{\mathbf{z}}$. 
A global rotation of this plane around $\hat{\mathbf{z}}$ is denoted by the azimuthal angle $\phi_{l}$. The hadronic part of the interaction is contained in the reaction plane within which the scattered nucleon has four momentum $P^{\mu}_{N} = (E_{N}, \mathbf{p_{N}})$ and the polar angle $\theta_{N}$ is defined between the vector $\mathbf{p_{N}}$ and $\hat{\mathbf{z}}$. The azimuthal angle between the reaction plane and the scattering plane is defined as $\phi_{N}$. The momentum transfer $Q^{\mu}$ is defined as 
\begin{equation}
Q^{\mu} = K^{\mu} - K^{\mu}_{l} = (\omega, \mathbf{q}).
\end{equation}

\noindent The missing energy of the system is defined as the energy that is passed to the residual nuclear system in the state it is in (excited or ground). This quantity is inherently dependent on the nuclear model used. In this model, it is defined as
\begin{equation}
E_{m} = \omega - T_{N} - T_{B},
\end{equation}
\noindent where $T_{N}$ is the scattered nucleon kinetic energy, $T_{B}$ is the kinetic energy of the recoil nucleus and $\omega$ is the energy transfer. The missing momentum is defined as 
\begin{equation}
\mathbf{p_{m}} = \mathbf{q} - \mathbf{p_{N}}.
\end{equation} 
\noindent For a given neutrino energy, the full neutrino-nucleus interaction is described by six independent variables; we choose the laboratory variables $(k_{l}, \theta_{l}, \phi_{l}, p_{N}, \theta_{N}, \phi_{N})$. With these independent variables, the fully exclusive cross section is given by
\begin{eqnarray}
\frac{d^{6}\sigma}{dk^{'}_{l} d\Omega_{k} dp_{N} d\Omega_{N}} &=& \frac{G_{F}^{2}\cos^{2}(\theta_{c})k^{2}_{l}p_{N}^{2}}{64 \pi^{5}} \frac{W_{B}}{E_{B} f_\text{rec}}\nonumber\\
&\times& L_{\mu\nu}\sum_\kappa\rho_\kappa(E_m) H_\kappa^{\mu\nu}.
\label{eq:full-d6sigma}    
\end{eqnarray}
\noindent Here, the cross section depends on the invariant mass of the residual nucleus, $W_{B}$, and the energy, $E_{B}$. The recoil factor, $f_\text{rec}$, is defined as 
\begin{equation}
f_\text{rec} = \Big| 1- \frac{\mathbf{p_{m}} \cdot \hat{\mathbf{z}}}{E_{B}} \Big|.
\end{equation}
\noindent The leptonic tensor is defined as 
\begin{equation*}
\begin{split}
L_{\mu\nu} =  \frac{2}{E_{\nu}E_{l}} \big[ K_{\mu}K_{l; \hspace{2pt}\nu} + K_{\nu} & K_{l; \hspace{2pt}\mu} - g_{\mu\nu} K^{\lambda}K_{l; \hspace{2pt} \lambda} \\
& - ih\epsilon_{\mu\nu\alpha\beta}K^{\alpha}K_{l}^{\beta} \big],
\end{split}
\end{equation*}
\noindent where $g_{\mu\nu}$ is the metric tensor (Minkowski metric), $\epsilon_{\mu\nu\alpha\beta}$ is the four-dimensional Levi-Civita tensor and $h$ is the helicity of the neutrino state, defined as $h=-1$ for neutrinos and $h=+1$ for antineutrinos. \\

\noindent $\rho_\kappa(E_{m})$ is the missing energy density for each shell $\kappa$. In the independent-particle shell model (IPSM), it is given by a Dirac delta centered at the binding energy of each shell $E_b^\kappa$, i.e.:
\begin{equation}
\rho_\kappa(E_m)=\delta(E_m-E_b^\kappa)\,.
\end{equation}

\noindent The hadronic tensor is more complex and encodes nuclear effects experienced by the scattered nucleon as it leaves the nuclear medium. It is defined through hadronic currents of single-particle RMF solutions as 
\begin{equation}
H^{\mu\nu}_{\kappa} = \frac{1}{2j+1}\sum_{m_{j}, s} [J^{\mu}_{\kappa, m_{j}, s}]^{*} [J^{\nu}_{\kappa, m_{j}, s}],
\end{equation}
where the sum is over the quantum numbers $m_{j}$ and $s$, which denote the third component of the total angular momentum of the hole state and spin of the scattered nucleon state, respectively. The quantum number $\kappa$ denotes the shell the scattered nucleon originated from. The quantum number $j$ is the total angular momentum of the shell and $2j+1$ gives the total occupancy available to the shell. The hadronic current is further defined through the four dimensional spinors for the bound nucleon state and scattered nucleon state, along with the transition operator defining the interaction:
\begin{equation}
J^{\mu}_{\kappa, m_{j}, s} 
= \int \text{d}\mathbf{p} \hspace{2pt} \bar{\psi}_s(\mathbf{p} + \mathbf{q}, \mathbf{p_{N}}) \hspace{2pt}  \Gamma^{\mu} \hspace{2pt}  \Psi_{\kappa}^{m_{j}}(\mathbf{p}).
\end{equation}
\noindent Here, $\psi$ is the scattered nucleon distorted wavefunction and $\Psi$ is the bound nucleon wavefunction. The transition operator, $\Gamma^{\mu}$, is that of the CCQE operator in the CC2 formalism~\cite{Kelly:PhysRevC.56.2672} and is defined as 
\begin{equation}
\label{CCQE_TO}
\Gamma^{\mu} = F_{1}\gamma^{\mu} + \frac{i F_{2}}{2M_{N}}\sigma^{\mu\nu}Q_{\nu} + G_{A}\gamma^{\mu}\gamma^{5} + \frac{G_{P}}{2M_{N}}Q^{\mu}\gamma^{5},
\end{equation}
\noindent where $F_{1}$, $F_{2}$, $G_{A}$ and $G_{P}$ are the vector, axial-vector and pseudoscalar form factors.  

\subsection{Initial state modelling}
\noindent The nuclear bound state is described with momentum distributions obtained within the RMF framework. RMF theory is an extension of the meson-mediated intranuclear force exchange model originally proposed by Walecka~\cite{Walecka-model} with extensions to add pseudovector mesons and nonlinear scalar meson self-couplings. While computationally intensive, the theory becomes more practical through a mean-field approximation in high-density nuclear regimes, where meson field operators are replaced by their expectation values. In this approach, the meson fields are treated classically, while the nucleon wavefunctions remain fully quantum mechanical ~\cite{computational-nuclear-physics-book}. This approximation renders the RMF model computationally tractable, allowing it to describe the ground state of medium- to high-density nuclei. \\

\noindent The Lagrangian density for this theory is defined as 
\begin{equation}
\begin{split}
\mathcal{L} &= \bar{\Psi}(i\gamma_{\mu}\partial^{\mu} - M)\Psi\\ 
&+ \frac{1}{2}\Big(\partial_{\mu}\sigma\partial^{\mu}\sigma - m_{\sigma}^{2}\sigma^{2}\Big) - U(\sigma) - g_{\sigma}\bar{\Psi}\sigma\Psi\\
& -\frac{1}{4}\Omega_{\mu\nu}\Omega^{\mu\nu} + \frac{1}{2}m_{\omega}^{2}\omega_{\mu}\omega^{\mu} - g_{\omega}\bar{\Psi}\gamma_{\mu}\omega^{\mu}\Psi\\
& -\frac{1}{4}\mathbf{R}_{\mu\nu}\mathbf{R}^{\mu\nu} + \frac{1}{2}m_{\rho}^{2}\boldsymbol{\mathbf{\rho}}_{\mu}\boldsymbol{\mathbf{\rho}}^{\mu} - g_{\rho}\bar{\Psi}\gamma_{\mu}\boldsymbol{\mathbf{\tau}}\boldsymbol{\mathbf{\rho}}^{\mu}\Psi\\
& -\frac{1}{4}F_{\mu\nu}F^{\mu\nu} - e\Big( \frac{1+\tau_{3}}{2} \Big)~\bar{\Psi}\gamma_{\mu}A^{\mu}\Psi + h.c.
\end{split}  
\end{equation}
\noindent Here, $U(\sigma) = \frac{1}{3}g_{2}\sigma^{3} + \frac{1}{4}g_{3}\sigma^{4}$ and mediates the non-linear self-coupling of the $\sigma$ meson. The field tensors for the meson fields are defined as:
\begin{equation}
\begin{split}
& \hspace{2pt} \Omega^{\mu\nu} = \partial^{\mu}\omega^{\nu} - \partial^{\nu}\omega^{\mu} \\
& \mathbf{R}^{\mu\nu} = \partial^{\mu}\boldsymbol{\mathbf{\rho}}^{\nu} - \partial^{\nu}\boldsymbol{\mathbf{\rho}}^{\mu} \\
& F^{\mu\nu} = \partial^{\mu}A^{\nu} - \partial^{\nu}A^{\mu}. 
\end{split}
\end{equation}

\noindent Such a Lagrangian has six free parameters: $g_{\sigma}$, $g_{2}$, $g_{3}$, $g_{\omega}$, $g_{\rho}$ and $m_{\sigma}$. Since the $\sigma$ meson is a phenomenological meson, $m_{\sigma}$ is a free parameter. The values chosen in this model is given below in Table~\ref{table:NLSH_RMF_parameters}.\\

\begin{table} [H]
    \centering
    \begin{tabular}{  c c c c c c  }
    %\hline \hline
    $m_{\sigma}$ & $g_{\sigma}$ & $g_{\rho}$ & $g_{\omega}$ & $g_{2}$ & $g_{3}$ \\ [0.5 ex]
    \hline \hline
    526.059 & 10.444 & 4.3830 & 12.945 & $-$6.9099 & $-$15.8337  \\ [1ex]
    %\hline
    \end{tabular}
    \caption{Parameters for the RMF model are chosen to match the NLSH set. All values are in MeV$/c$~\cite{NLSH}.}
    \label{table:NLSH_RMF_parameters}
\end{table}

\noindent The mean field approach allows the treatment of the meson fields to be classical and so when the Euler-Lagrange equation is applied, classical Klein-Gordon fields are obtained. For nucleons, this produces the Dirac equation. The nuclear bound states are the stationary state solutions to these equations. The Dirac equation for the bound nucleon state is 
\begin{equation}
\label{eq:RMF-dirac-eq-matrix}
\Big[ -i \boldsymbol{\mathbf{\alpha}}\cdot\boldsymbol{\mathbf{\nabla}} + g_{V} V^{0}(r) + \beta\big(M - g_{S}S(r)\big) \Big]\Psi(\mathbf{r}) = E \hspace{2pt} \Psi(\mathbf{r}),
\end{equation}
\noindent where the vector and scalar fields have been combined to form $S(r)$ and $V^{0}(r)$ and have respective couplings $g_{S}$ and $g_{V}$ for brevity. Here, we consider only spherically symmetric nuclei; therefore, the meson fields depend solely on the radial coordinate, $r$. Solutions can be found by expanding in terms of positive energy solutions and creation operators:
\begin{equation}
\Psi(\mathbf{r}) = \sum_{q} A_{q}\mathcal{U}_{q}(\mathbf{r}),
\end{equation}
\noindent where $q$ is the set of quantum numbers corresponding to the state. It can be explicitly expressed as $q = \{n, l, j, m, t\}$. $\mathcal{U}_{q}$ are the positive energy solutions and $A_{q}$ are the baryon creation operators. We seek solutions for $\mathcal{U}_{q}$ in the form of the following ansatz:
\begin{equation}
\mathcal{U}_{q}(\mathbf{r}) = \begin{pmatrix}i \frac{G_{\alpha}(r)}{r}\Phi_{\kappa m} \\ - \frac{F_{\alpha}(r)}{r}\Phi_{-\kappa m} \end{pmatrix}\zeta_{t}.
\end{equation}

\noindent There are two wavefunctions, $G(r)$ and $F(r)$ that correspond to the upper and lower spinor component wavefunctions. $\alpha$ is given as the quantum numbers $n, \kappa, t$. Spherical spinors $\Psi$ contain the angular dependence. $\zeta$ is the isospin projection operator where $t = 1/2$ is for protons and $t=-1/2$ is for neutrons. The nucleon wavefunctions satisfy the following coupled differential equations:
\begin{align}
\label{eq:RMF-coupled-ground-state-eqs1}
\begin{split}
& \frac{\partial F_{\alpha}(r)}{\partial r} - \frac{F_{\alpha}(r)}{r}\kappa \\ 
& +  \big[ E_{\alpha} - g_{V}V^{0}(r) - M + g_{S}S(r) \big] G_{\alpha}(r) = 0 
\end{split}
\end{align}

\begin{align}
\label{eq:RMF-coupled-ground-state-eqs2}
\begin{split}
& \frac{\partial G_{\alpha}(r)}{\partial r} + \frac{G_{\alpha}(r)}{r}\kappa \\
&- \big[ E_{\alpha} - g_{V}V^{0}(r) + M + g_{S}S(r) \big] F_{\alpha}(r) = 0.  
\end{split}    
\end{align}

\noindent The meson field densities given by $\rho_{S}$ and $\rho_{B}$ are
\begin{equation}
\label{eq:scalar-densities}
\rho_{S}(r) = \sum_{i}^{\text{occ}} \frac{(2j+1)}{4\pi r^{2}} \big( |G(r)|^{2} - |F(r)|^{2} \big), 
\end{equation}
\begin{equation}
\label{eq:vector-densities}
\rho_{B}(r) = \sum_{i}^{\text{occ}}   \frac{(2j+1)}{4\pi r^{2}} \big( |G(r)|^{2} + |F(r)|^{2} \big),
\end{equation}
\noindent where the summation is over the occupied states only due to the use of the no-sea approximation. It is important to note that since most neutrino detectors are constructed from nuclei that are spherically symmetric or nearly so, such as $\ce{^{16}O}$, $\ce{^{40}Ar}$, and $\ce{^{12}C}$, the RMF model, with its approximations, is particularly well-suited for simulating neutrino interactions in these nuclei.

\subsubsection{Numerical solutions to the RMF equations}
\noindent The RMF equations can be solved in a Dirac-Hartree approximation using a self-consistent method. The algorithm for this is given in Ref.~\cite{computational-nuclear-physics-book} and uses software called \texttt{TIMORA}, modified to include the non-linear $\sigma$ meson self-couplings. The algorithm is as follows:

\begin{enumerate}
\item Propose a guess for the initial potentials $S(r)$ and $V^{0}(r)$. A suitable guess uses a Woods-Saxon potential. 
\item Solve the Dirac equations (Eq.~\ref{eq:RMF-coupled-ground-state-eqs1} and~\ref{eq:RMF-coupled-ground-state-eqs2}) to obtain the upper and lower spinor component wavefunctions $G(r)$ and $F(r)$.
\item From these wavefunctions, calculate the meson field densities $\rho_{S}(r)$ and $\rho_{B}(r)$ using Eq.~\ref{eq:scalar-densities}~and~\ref{eq:vector-densities}. 
\item Solve the meson field Klein-Gordan equations to obtain new expressions for the potentials.
\item Substitute these new potentials back into Step 1 and repeat until convergence.
\end{enumerate}

\noindent The convergence requirement is enforced by comparing the change in energy eigenvalue $E$ between iterations to a convergence requirement input. That is, the change in the energy eigenvalues between the current iteration and the previous must be less than the convergence requirement, for all nucleons. 

\subsection{Final state modelling}
\noindent The scattered nucleon wavefunction is the solution to the Dirac equation in the presence of a relativistic nuclear potential. The choice of nuclear potential alters the scattered nucleon wavefunctions. 
If there is no potential, then the solution is that of a plane wave. If there is a potential, then the dispersion relation for the nucleon is altered, leading to the distorted wave approach. In what follows, we describe the final-state nuclear potentials used in this work.

\subsubsection{RMF and energy-dependent RMF}

\noindent The same RMF potential employed to compute the momentum distributions for the initial state nucleons, can be used to produce solutions for the scattered state. This leads to the same Dirac equation for both the ground state and the scattered state and, if the bound and scattered state wavefunctions belong to an orthonormal basis, the Pauli exclusion principle is naturally implemented as described in Refs.~\cite{Nikolakopoulos19,Gonzalez-Jimenez19}.
However, the RMF potential is energy-independent. As shown in Ref.~\cite{Gonzalez-Jimenez20}, at high scattered nucleon momenta, these potentials are too strong and the QE peak is shifted towards $\omega$ values that are too high compared to inclusive electron scattering data. To address this issue, the energy-dependent RMF potential (EDRMF) was developed in Ref.~\cite{Gonzalez-Jimenez19} by modifying the RMF potential with a blending function that varies with the scattered nucleon kinetic energy. The parameters in this function are taken from SuSAv2 scaling parameters~\cite{Gonzalez-Jimenez14, Megias16}, which are tuned to fit electron-scattering data. This results in the EDRMF potential getting attenuated at large nucleon energies, in a similar fashion to the real part of phenomenological optical potentials and more consistent with dispersion relationships~\cite{Gonzalez-Jimenez20}. \\

\noindent The EDRMF potential is a real potential by construction. As a result, there is no flux lost to the inelastic channel. The results obtained with this approach are extremely similar to those obtained when one uses the real part of an energy-dependent optical potential (fitted to elastic proton-nucleus scattering data)~\cite{Gonzalez-Jimenez20} or within a relativistic Green function approach~\cite{Meucci03}. It has also been previously discussed that the use of real potential together with the NEUT cascade does not double count any effects due to FSI~\cite{Nikolakopoulos22}. The NEUT cascade does not include any elastic FSI channels; however, if a different generator does include elastic FSI channels, then they must be turned off.
For example, the GiBUU generator uses transport equations to calculate final-state particle trajectories rather than a cascade model; such a framework may lead to difficulties in decoupling double counting effects.

\subsubsection{Relativistic Optical Potential (ROP)}
\noindent ROPs have been extracted from proton-nucleus elastic scattering data~\cite{Cooper-potentials} and are widely used in the field~\cite{Udias93, Giusti11, Butkevich12, Gil21}. Such potentials are complex, meaning they contain a real and imaginary component. The idea of optical potentials comes from the refraction of light where imaginary components are introduced to account for absorption within the medium. The imaginary component can be used to account for such ``loss'' in a system, while the real part accounts for the elastic interactions. \\

\noindent The energy of the scattered proton is redistributed by the real component of the potential in a similar fashion to how the Fermi factor affects the electron wavefunction in beta decay processes.
This causes a redistribution of the strength of the predicted cross sections. If one uses the real component only, the nuclear effects probed by proton scattering are retained while not applying any reduction in flux to inelastic channels. This approach matches EDRMF potentials well at high scattered nucleon momentum but differs when the momentum is lower, due to the lack of orthogonality between the initial and final states in the ROP approach~\cite{Gonzalez-Jimenez20,Gonzalez-Jimenez19}.\\

\noindent In this work, we also show predictions with the full complex ROP. In particular, the energy-dependent $A$-independent potential fit to elastic proton-nucleus scattering data, called the EDAI potential~\cite{Cooper-potentials}, is used. For interactions on carbon, the proton-$^{12}$C elastic scattering data is used and the model is referred to as the EDAIC model. 
For interactions on $^{40}$Ar, there is no optical potential available; therefore, the potential for the isobaric $^{40}$Ca is used and is denoted as EDAICa in this work. We point out that the Coulomb potential for calcium ($Z=20$) is added analytically in the fits of Ref.~\cite{Cooper-potentials}, therefore, we add it correctly for argon ($Z=18$).\\

\noindent The predictions from this model correspond to a scenario where the struck nucleon exits the nucleus without losing energy, except for that lost to recoil in elastic interactions.
This should be equivalent to the output from a cascade when only the nucleons that pass through the nucleus without interacting are retained. In this case, the CCQE final state would consist of a muon and exactly one proton, so the EDAI model provides CCQE predictions that can be considered as a lower bound when compared with the semi-inclusive neutrino-nucleus scattering samples that are analysed in Section~\ref{sec:resutls}.

\subsection{Beyond the independent-particle shell model}

\noindent The energy and momentum distributions of nucleons inside the nucleus are not fully described within the IPSM, as correlations are present inside the nucleus. IPSM models have been shown to overestimate data indicating that shell occupancies in nature are lower compared to IPSM predictions~\cite{Garino92, Holtrop98, Dutta03, Jiang22, Jiang23, Fissum04, Kramer89, Yasuda10, Volkov90, Udias93, Giusti11, Atkinson18}.
A more realistic model including correlation effects is necessary. Correlations can be incorporated through a spectral function approach which is computed from a nuclear model that explicitly includes them.
The spectral function contains a shell-model contribution, where the energy distribution and occupations of the nucleons appear smeared and depleted, with respect to the IPSM prediction. An additional component coming from nuclear correlations, which causes nucleons to appear at higher energy, is also included~\cite{Benhar08}. \\

\noindent Spectral function calculations to nucleon knock-out assume a factorisation of the cross section into the ``spectral-function'' term, that is the probability of finding a nucleon of a given energy and momentum, and the free lepton-nucleon cross section. This may not be a good approximation in every circumstance, for instance, it provides the wrong ratio for electron-neutrino and muon-neutrino charged current QE cross sections~\cite{Nikolakopoulos19} in an important part of the phase space. However, it is possible to build a representation of a realistic spectral function, where the shell-model content of it is written as the sum of momentum distributions computed within the RMF, plus a term describing the high-energy high-momentum tail~\cite{Gonzalez-Jimenez22}.
\\

\noindent Thus,  in the model used in this work, Gaussian functions with parameters fit to replicate the Rome SF~\cite{Benhar94,Benhar05} missing energy density are used for carbon and oxygen, instead of Dirac delta functions. The central values of the Gaussian functions are described by the RMF eigenvalues, or experimental values when they are available, and the widths are now finite~\cite{Gonzalez-Jimenez22, Franco-Patino22}. Following the approach of~\cite{Gonzalez-Jimenez22, Franco-Patino22, Franco-Munoz23, Franco-Patino24}, an additional shell is introduced in order to capture the strength that appears at high missing energy and high missing momentum due to short-range correlations~\cite{Egiyan06, Duer18}. \\

\noindent The final missing energy density is given as a sum over the densities of each shell:
\begin{equation}
\rho(E_{m}) = \sum_{\kappa}\rho_{\kappa}(E_{m}),
\end{equation} 
with $\int dE_m \rho(E_m)=Z$ or $N$, the total number of protons or neutrons.
The details of the Gaussian parameters and missing energy densities are given in Appendix~\ref{sec:app:A}. 

\noindent We point out that, as shown in Eq.~\ref{eq:full-d6sigma}, the function $\rho(E_m)$ is independent of the hadronic tensor and is implemented analytically in the code. In contrast, the hadronic tensors are precomputed and stored in tables (see next section). Consequently, the parameters defining $\rho(E_m)$ can be modified by NEUT users. However, the chosen configuration should always be verified for compatibility with electron scattering data. For the well-studied $\ce{^{12}C}$ nucleus, the available electron-scattering data constrains the shell densities to a large extend and thus there is minimal flexibility. However, for heavier and more complex nuclei, like $\ce{^{40}Ar}$ and $\ce{^{40}Ca}$, the parameter uncertainties are significantly larger~\cite{Franco-Patino24, Franco-Munoz25}.

\section{Implementation in the NEUT generator}
\label{sec:implementation}
\noindent This section details the implementation of the model into the NEUT neutrino event generator framework. The model, due to its computational complexity, is implemented using precomputed hadronic tensors in a tabulated form. There are precomputed hadronic tensor tables for each neutrino helicity state (neutrino and antineutrino) and for each nuclei. Currently, the tables for $^{12}$C, $^{16}$O and $^{40}$Ar are implemented. There are also tables for each nuclear potential. The hadronic tensor values can be read directly from the table and contracted with the leptonic tensor without any further integration. The hadronic tensor tables were generated with a dipole axial form factor with axial mass $M_{A}^{QE} = 1.05$~GeV. A table of the total cross section and maximum six-fold differential cross section as a function of neutrino energy are also required. These two sets of tables are precomputed for each nuclei, nuclear potential and neutrino helicity and are read in by NEUT. The maximum six-fold differential cross section is used as a ceiling of the accept-reject algorithm. The value of the ceiling is increased by 10\% to compensate for finite MC statistics leading to an underestimation of the ceiling.  \\

\noindent An improved efficiency was achieved by the introduction of physical energy-dependent constraints on the phase spaces used for the rejection method when generating events. The ceiling of the accept-reject algorithm was also changed to be thrown uniformly from $0$ to a neutrino energy-dependent maximum; this minimised the number of reject events. The largest improvement in speed came from changing the thrown kinematic variables by use of a Jacobian and draw the missing energy from the pre-defined distribution instead of throwing it uniformly. The details and structure of the hadronic tensor tables and interpolation are given in Appendix~\ref{sec:app:B} and further details for the implementation are given in Appendix~\ref{sec:app:D}.

\subsection{Initial algorithm and improvements}
The initial algorithm was as follows.
\begin{enumerate}
	\item Uniformly throw the six independent variables to describe the scattered state system given an incident neutrino energy.
	\item Calculate $E_{m}$, $p_{m}$, $Q^{2}$ and $\omega$.
	\item Reject any events that have such values outside of theoretically allowed bounds for these quantities.
	\item Rotate to get to frame of reference of the hadronic tensor tables.
	\item Interpolate tensor values and rotate back to lab frame. 
	\item Calculate the six-fold differential cross section.
	\item Accept or reject event based on random throw of the six-fold differential cross section. The ceiling is an input.
	
\end{enumerate}
In the above algorithm, it was found that it was particularly slow due to the bound check on $E_{m}$. The missing energy distribution is very peaked and thus a general MC accept-reject algorithm would be inefficient in reproducing it.
As a result, the original algorithm was changed such that $E_{m}$ was drawn directly from the true distribution using it as a probability density function. This required an additional weighting of the final accept-reject ceiling by a factor of $\frac{1}{\rho(E_{m})}$. 
A respective Jacobian was also required in order to obtain the six-fold differential cross section given in Eq.~\ref{eq:full-d6sigma} from the following differential cross section:
\begin{equation}
\label{eq:full-d6sigma-Em}
\frac{d^{6}\sigma}{d\mathbf{k_{l}} dE_{m} d\Omega_{N} }.
\end{equation}
\noindent The Jacobian that is used neglects nuclear recoil and is given in Eq.~\ref{eq:jacobian} as
\begin{equation}\label{eq:jacobian}
\Bigg | \frac{\partial p_{N}}{\partial E_{m}} \Bigg | = \frac{E_{N}}{p_{N}}.
\end{equation}

\noindent The effect of nuclear recoil was found to be sub-percent level. The details of the derivation for both the Jacobian neglecting and including nuclear recoil are given in Appendix~\ref{sec:app:C}.

\subsection{Final algorithm}
\noindent The final algorithm is given as 
\begin{enumerate}
	\item Uniformly throw the five independent variables to describe the scattered state system given an incident neutrino energy.
	\item Sample $E_{m}$ from missing energy density.
	\item Calculate $p_{m}$, $Q^{2}$ and $\omega$.
	\item Reject any events that have such values outside of theoretically allowed bounds.
	\item Transform to reference frame of hadronic tensor tables.
	\item Interpolate tensor values and rotate back to lab frame. 
	\item Calculate the six-fold differential cross section.
	\item Accept or reject event based on random throw of the six-fold differential cross section. The ceiling is an input.
\end{enumerate}

\section{Results}
\label{sec:resutls}
\begin{figure}[H]
        \includegraphics[keepaspectratio, width=0.5\textwidth]{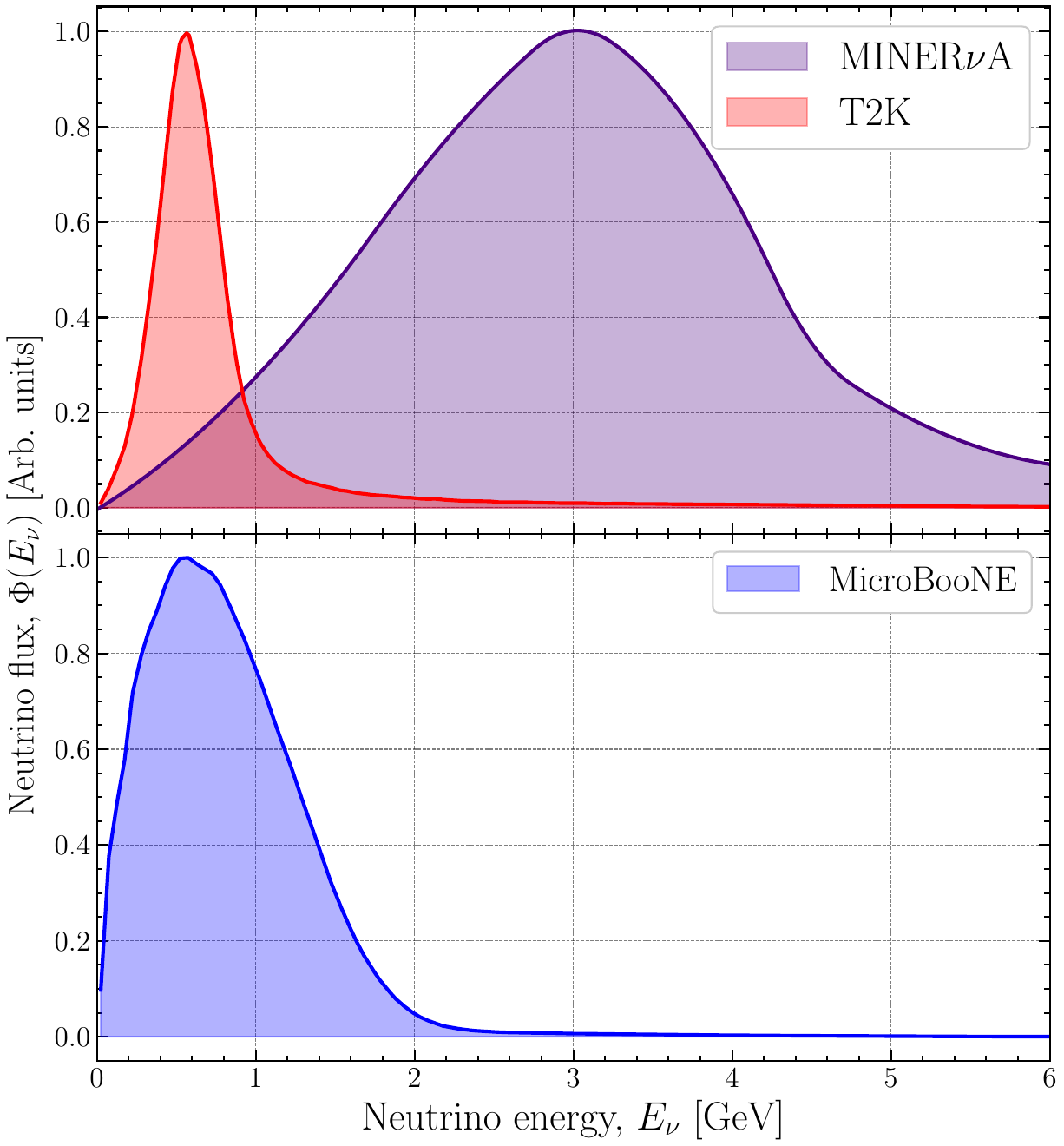}
        \caption{Comparison of the neutrino fluxes for the T2K, MINER$\nu$A and MicroBooNE experiments. The fluxes are peak normalised for shape comparison. The T2K and MINER$\nu$A (\textit{top}) target is hydrocarbon while MicroBooNE (\textit{bottom}) is on argon.}
        \label{fig:exp-fluxes}
\end{figure}

\noindent In this section, we benchmark the model on hydrocarbon and argon using cross-section measurements from the T2K~\cite{T2K_2018_data}, MINER$\nu$A~\cite{MinervaData:2018, minerva-data-set}, and MicroBooNE experiments~\cite{uboone-kinematics:2020, uboone-TKI:2023, uboone-TKI:2023:long}. The respective experimental neutrino fluxes are in Fig.~\ref{fig:exp-fluxes} and show that MicroBooNE and T2K have a similar average peak neutrino energy, so the primary interaction channel would be CCQE for both. Data comparisons were performed using the NUISANCE framework~\cite{Nuisance}. All datasets contain information on the final-state protons and report transverse kinematic imbalance (TKI) cross sections and respective covariance matrices. These variables are more sensitive to the initial-state nuclear modelling and better quantify the nuclear effects. The definition of TKI variables are:
\begin{equation}
\label{eq:tki}
    \begin{split}
        & \delta p_{T} = |\mathbf{p}_{T}^{~l}~+~ \mathbf{p}_{T}^{~p} |, \\
        & \delta \alpha_{T} = \arccos \Big(\frac{-\mathbf{p_{T}}^{l} ~\cdot ~\delta \mathbf{p_{T}}}{p_{T}^{l} \delta p_{T}^{l}} \Big), \\
        & \delta \phi_{T} = \arccos \Big( \frac{-\mathbf{p_{T}}^{l} ~\cdot ~\mathbf{p_{T}}^{p}}{p_{T}^{l} p_{T}^{p}} \Big).
    \end{split}
\end{equation}

\noindent Here, $p^{l} $ and $p^{p}$ represent the lepton and leading proton momentum, respectively, and $T$ indicates the plane transverse to the direction of the neutrino beam. \\

\noindent The T2K measurement is presented in the unfolded truth space using a binned likelihood fit and a standard regularisation procedure~\cite{T2K_2018_data}. The full covariance matrices are released with the dataset.
The MINER$\nu$A measurement, like T2K, is presented in the unfolded truth space using an unfolding technique based on Bayes' theorem~\cite{minerva-data-set, DAGOSTINI1995487}. Full covariance matrices are released alongside the dataset. The MicroBooNE measurements in Refs.~\cite{uboone-TKI:2023,uboone-TKI:2023:long} use an unfolding procedure based on Ref.~\cite{Tang_2017}, while Ref.~\cite{uboone-kinematics:2020} uses forward folding. Both procedures produce an additional smearing matrix that is used to apply detector smearing affects in the case of the forward folding approach, and contains information on the amount of bias and regularisation in the unfolding approach. Since this smearing matrix is applied to the published unfolded cross sections, it must also be applied to any theoretical predictions being compared to the measured cross sections~\cite{uboone-TKI:2023:long}. The full covariance matrices and smearing matrices are released alongside the dataset. \\

\noindent We report $\chi^{2}$ analysis along with the total number of degrees of freedom in order to decide whether a model prediction is in agreement with a measurement. The $\chi^{2}$ definition used in this work is given by

\begin{equation*}
    \chi^{2} = \sum_{i,~j} \big( D - M \big)_{i} \big(\text{Cov}^{-1} \big)_{ij} \big( D - M\big)_{j}.
\end{equation*}

\noindent Here, $i$ and $j$ are matrix indices that indicate a value in a particular bin of a measurement. $D$ and $M$ represent the data and MC value in a given bin respectively. The inverse covariance matrix is given by $\text{Cov}^{-1}$ and has dimensions $i \times j$. \\

\noindent The NEUT framework is used to benchmark the EDRMF model against other models implemented within NEUT. The implementation of the spectral function, based on the Rome spectral function~\cite{Ankowski15}, is indicated as ``SF'' and was generated with an axial mass of $M_{A}^{QE} = 1.03$~GeV. There is currently no argon spectral function implemented in NEUT. The model presented in Ref.~\cite{nieves-model}, referred to as ``N1p1h'', was generated using $M_{A}^{QE} = 1.05$~GeV. NEUT 6.0.0 was used for all samples except the N1p1h model, for which NEUT 5.8.0 was used. \\

\noindent The NEUT cascade is applied to all samples unless otherwise stated with ``no cas''. All samples were produced using the NEUT default run mode meaning that 2p2h, pion production and deep inelastic scattering channels are produced at a rate that is proportional to their total cross sections. In NEUT, the 2p2h channel is based on Refs.~\cite{Nieves11, nieves-model} and the pion production model is the Rein-Seghal model with added lepton mass corrections~\cite{Rein:1980wg, PhysRevD.76.113004, LepMassEff-Graczyk-Sobczyk-PhysRevD.77.053003}. The joint contribution of 2p2h and $\pi$-absorption is shown separately with the cascade applied. The 2p2h contribution without the cascade is also shown separately. $\pi$-absorption occurs in the cascade so there is no such contribution if the cascade is off. \\

\noindent For the EDRMF model, the scattered nucleon is inserted into the nucleus at a uniform random position, which is then propagated through the cascade model. This approach contrasts with using a more realistic nuclear density distribution, where the scattered nucleon is inserted at a position which reflects the spatial probability of the underlying nuclear shell structure. Consequently, while uniform sampling simplifies the approach, it neglects important details of the nuclear structure. 

\subsection{T2K data comparison}
\label{sec:T2K}
\noindent CC$0\pi$N$p$ (``semi-inclusive'') datasets~\cite{T2K_2018_data} are used to benchmark the model on hydrocarbon. The applied kinematic selections are in given in Table~\ref{table:T2K_2018_data_cuts}. The dataset is a triple-differential cross section from which double-differential and single-differential results are extracted. As a result, a $\chi^{2}$ value is available only for the full triple-differential cross section and the TKI results. \\

\begin{table} [H]
    \centering
    \begin{tabular}{ c | c c }
    \hline \hline
    Kinematic Variable & CC0$\pi$Np & TKI \\ [0.5 ex]
    \hline \hline
    $p_{p}$ & $>$ 500 MeV & 0.45 -- 1 GeV \\
    $\cos \theta_{p}$ & $\dots$ & $>$ 0.4 \\
    $p_{\mu}$ & $\dots$ & $>$ 250 MeV \\
    $\cos \theta_{\mu}$ & $\dots$ & $>$ $-$0.6 \\ [1ex]
    \hline
    \end{tabular}
    \caption{Kinematic selections applied in the T2K dataset. Here ``$\dots$'' indicate that there is no selection applied. In the T2K analysis, any number of protons are selected within the kinematic selections but only the highest momentum proton is used in the TKI and proton cross sections.}
    \label{table:T2K_2018_data_cuts}
\end{table}

\subsubsection{Single-differential sample}
\begin{figure}
        \includegraphics[keepaspectratio, width=0.45\textwidth]{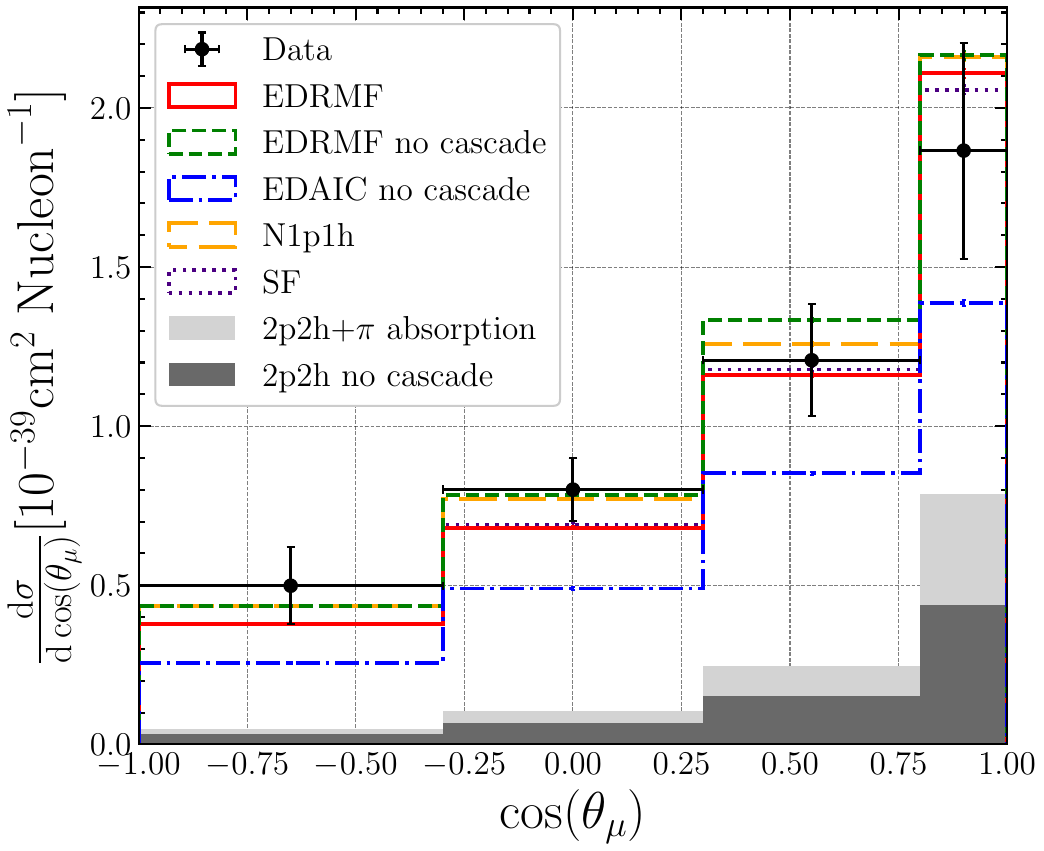}
        \caption{T2K $\nu_{\mu}$ CC$0\pi$N$p$ dataset on hydrocarbon. Single differential cross section given as a function of scattered lepton angle. The implemented EDRMF model with the cascade applied is represented by the solid red line, while the version without the cascade is shown as the dashed green line. The EDAIC model without the cascade is depicted by the blue dash-dotted line. The N1p1h model is indicated by the long dashed yellow line, and the SF model is shown as the purple dotted line. The 2p2h and $\pi$-absorption contributions are represented by the light gray filled bar, whereas the 2p2h-only contribution without the cascade is depicted by the dark gray filled bar. Data taken from~\cite{T2K_2018_data}.
        }
        \label{fig:t2k2018:1p_costheta}
\end{figure}

\noindent The single-differential cross section is given in Fig.~\ref{fig:t2k2018:1p_costheta}. In the low $\cos(\theta_{\mu})$ region, where the 2p2h and $\pi$-absorption contributions are low, the EDRMF model with the cascade applied predicts values slightly below the data but remains within the error bars. In contrast, in the high $\cos(\theta_{\mu})$ region, the model predicts values above the data, though still consistent within the error margins. In this region, the contributions from 2p2h and $\pi$-absorption are significantly higher, making it challenging to disentangle whether the observed excess originates from CCQE or non-CCQE contributions.\\

\noindent The $\chi^{2}$ values for the full multi-differential sample are presented in Table~\ref{table:mutlidiff_chi2}. Interestingly, the EDAIC model achieves the best agreement with the data based on the $\chi^{2}$ evaluation.

\begin{table} [H]
    \centering
    \begin{tabular}{  c |  c   }
    \hline \hline
    Model & $\chi^{2}/N_{dof}$ \\ [0.5 ex]
    \hline \hline
    EDRMF cas & $429.9/93$  \\
    EDRMF no cas & $471.9/93$  \\
    N1p1h cas & $336.4/93$  \\
    SF cas & $490.3/93$  \\
    EDAIC no cas & $259.7/93$  \\ [1ex]
    \hline
    \end{tabular}
    \caption{The T2K dataset $\chi^{2}$ values for the triple-differential cross section. ``cas'' and ``no cas'' indicate where the cascade has and has not been applied respectively.
    }
    \label{table:mutlidiff_chi2}
\end{table}

\subsubsection{Double- and triple-differential samples}
\begin{figure*}
    \begin{minipage}[h]{\textwidth}
    \includegraphics[keepaspectratio, width=0.85\textwidth]{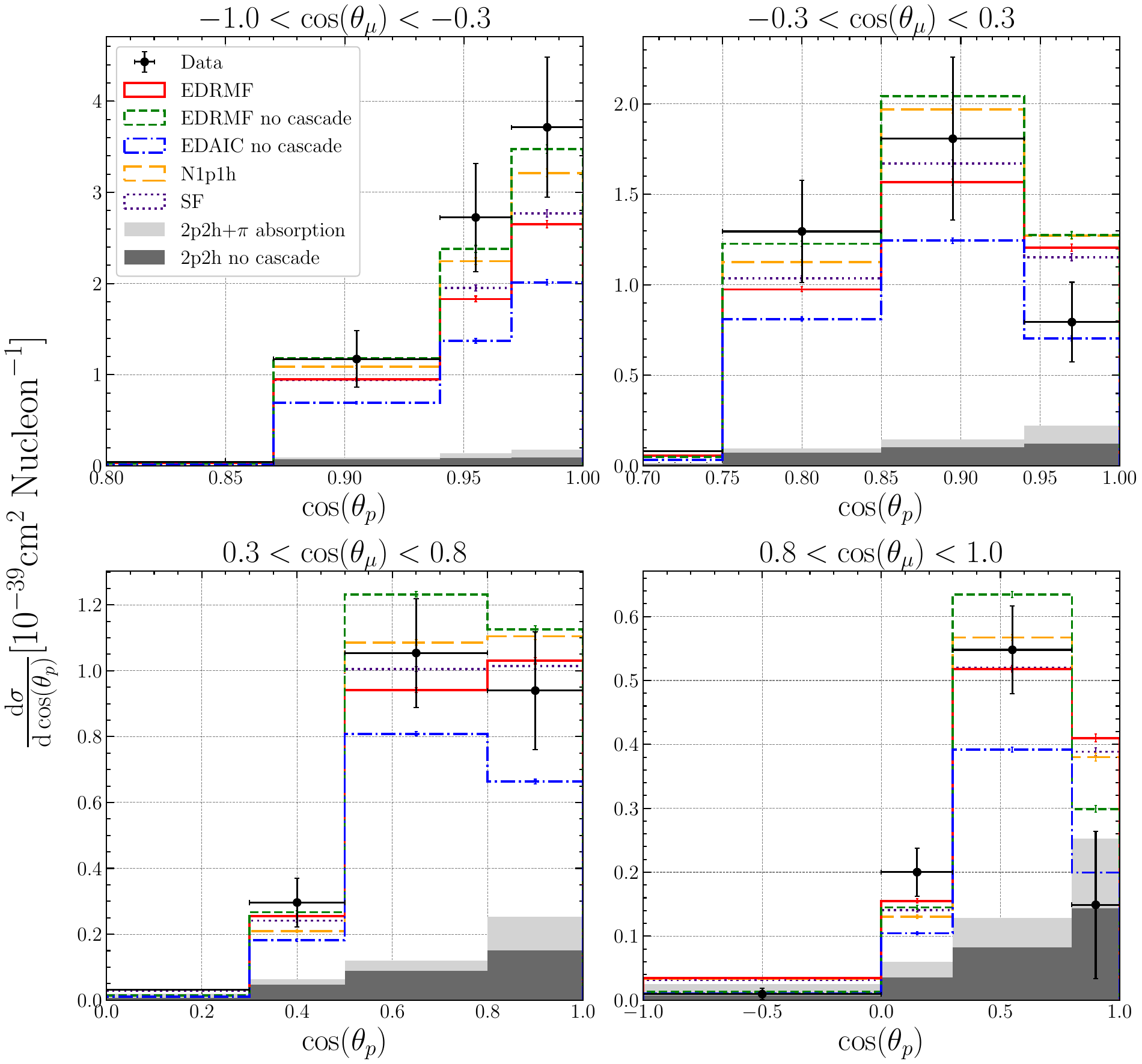}
    \caption{T2K $\nu_{\mu}$ CC$0\pi$N$p$ dataset on hydrocarbon. The differential cross section is given as a function of the scattered proton angle in slices of the scattered lepton angle starting from the top left. The histograms follow the same definition as Fig.~\ref{fig:t2k2018:1p_costheta}. Data taken from~\cite{T2K_2018_data}.}
    \label{fig:t2k2018:1p_costhetap_slices}
    \end{minipage}
\end{figure*}

\noindent Fig.~\ref{fig:t2k2018:1p_costhetap_slices} shows the differential cross section as a function of the leading proton scattering angle in slices of the lepton scattering angle. For low $\cos(\theta_{\mu})$, there is little contribution from 2p2h and $\pi$-absorption. In these regions, the EDRMF model with the cascade applied underestimates the data at high $\cos(\theta_{p})$. The EDRMF with cascade applied is also the lowest when compared to SF and N1p1h models. In the last bin in $-0.3 < \cos(\theta_{\mu}) < 0.3$, the EDRMF model overestimates the data along with other models. In this region the 2p2h and $\pi$-absorption contributions are high and all models overestimate the data suggesting that the 2p2h and $\pi$-absorption may be too high in NEUT. \\

\noindent 
As $\cos(\theta_{\mu})$ increases, all models within NEUT begin to overestimate the data, with a significant overestimate in the last bin of the $0.8 < \cos(\theta_{\mu}) < 1.0$ sample. 
In this bin, the `2p2h+$\pi$-absorption' contribution alone is quite higher than the center of the experimental value,
suggesting that the NEUT 2p2h and $\pi$-absorption contributions are given too much strength at these kinematics. \\

\noindent
The EDAIC model, which represents the case where the struck nucleon experiences only elastic FSI, is expected to provide a lower bound for the QE channel. As anticipated, the EDAIC model consistently underestimates the data.\\

\begin{figure*}
    \begin{minipage}[h]{\textwidth}
    \includegraphics[keepaspectratio, width=0.85\textwidth]{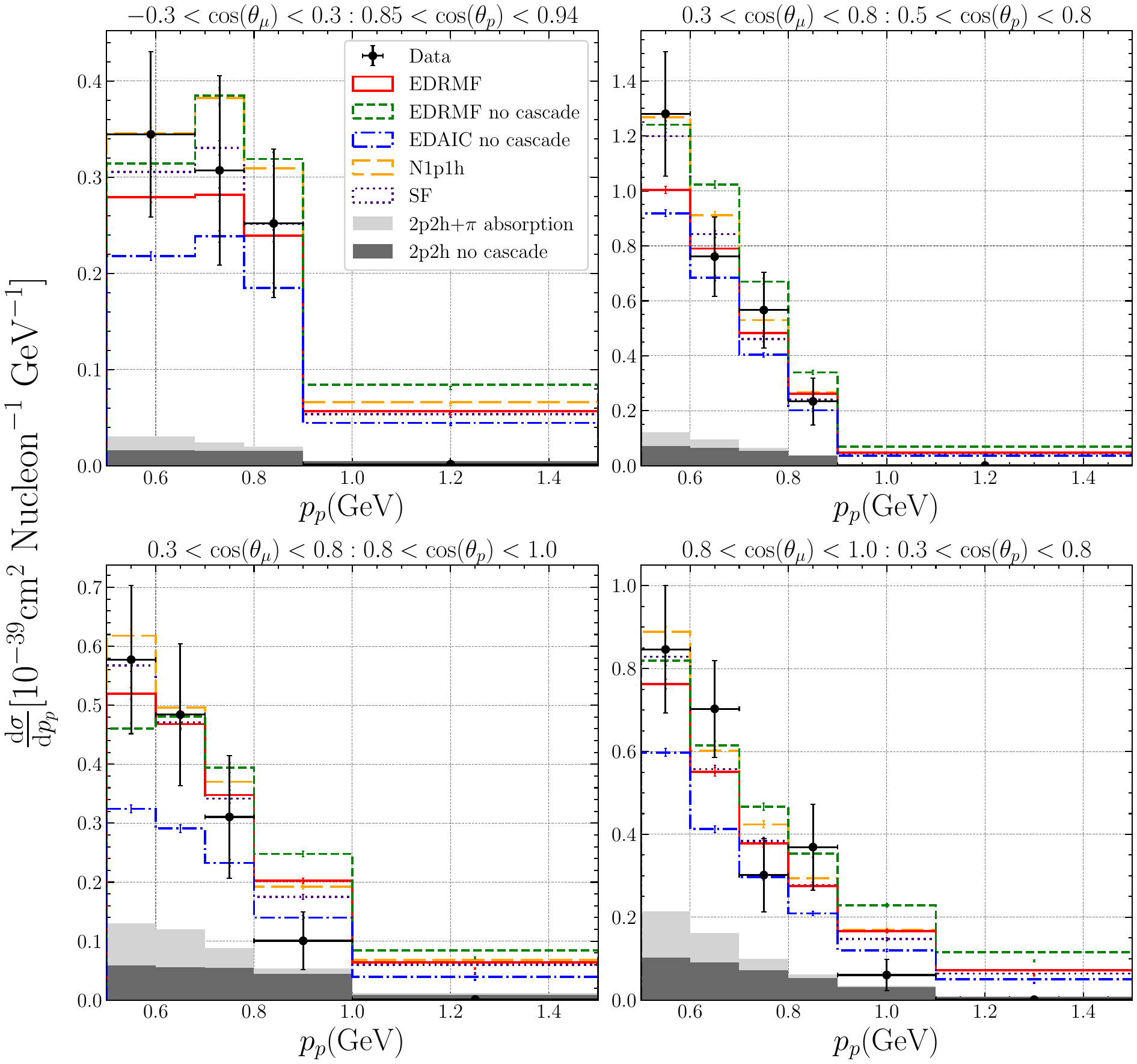}
    \caption{T2K $\nu_{\mu}$ CC$0\pi$N$p$ dataset on hydrocarbon. The differential cross section is given as a function of the scattered proton momentum in slices of the scattered lepton and nucleon angle starting from the top left. The histograms follow the same definition as Fig.~\ref{fig:t2k2018:1p_costheta}. Data taken from~\cite{T2K_2018_data}.}
    \label{fig:t2k2018:1p_costhetap_pp_slices}
    \end{minipage}
\end{figure*}

\noindent
Fig.~\ref{fig:t2k2018:1p_costhetap_pp_slices} shows the differential cross-section as a function of the leading proton momentum, divided into slices based on the scattering angles of the leading proton and lepton. For $p_{p} > 1.0$ GeV, all models overestimate the data, despite negligible contributions from 2p2h and $\pi$-absorption processes. However, this region contains very few events in the dataset, making it statistically limited. It is interesting to note that in the $-0.3 < \cos(\theta_{\mu}) < 0.3$, $0.85 < \cos(\theta_{p}) < 0.94$ slice, for $p_{p} < 1.0$ GeV, no model replicates the shape seen in data. 

\subsubsection{TKI samples}
\begin{figure*}
    \begin{minipage}[h]{\textwidth}
    \includegraphics[keepaspectratio, width=0.85\textwidth]{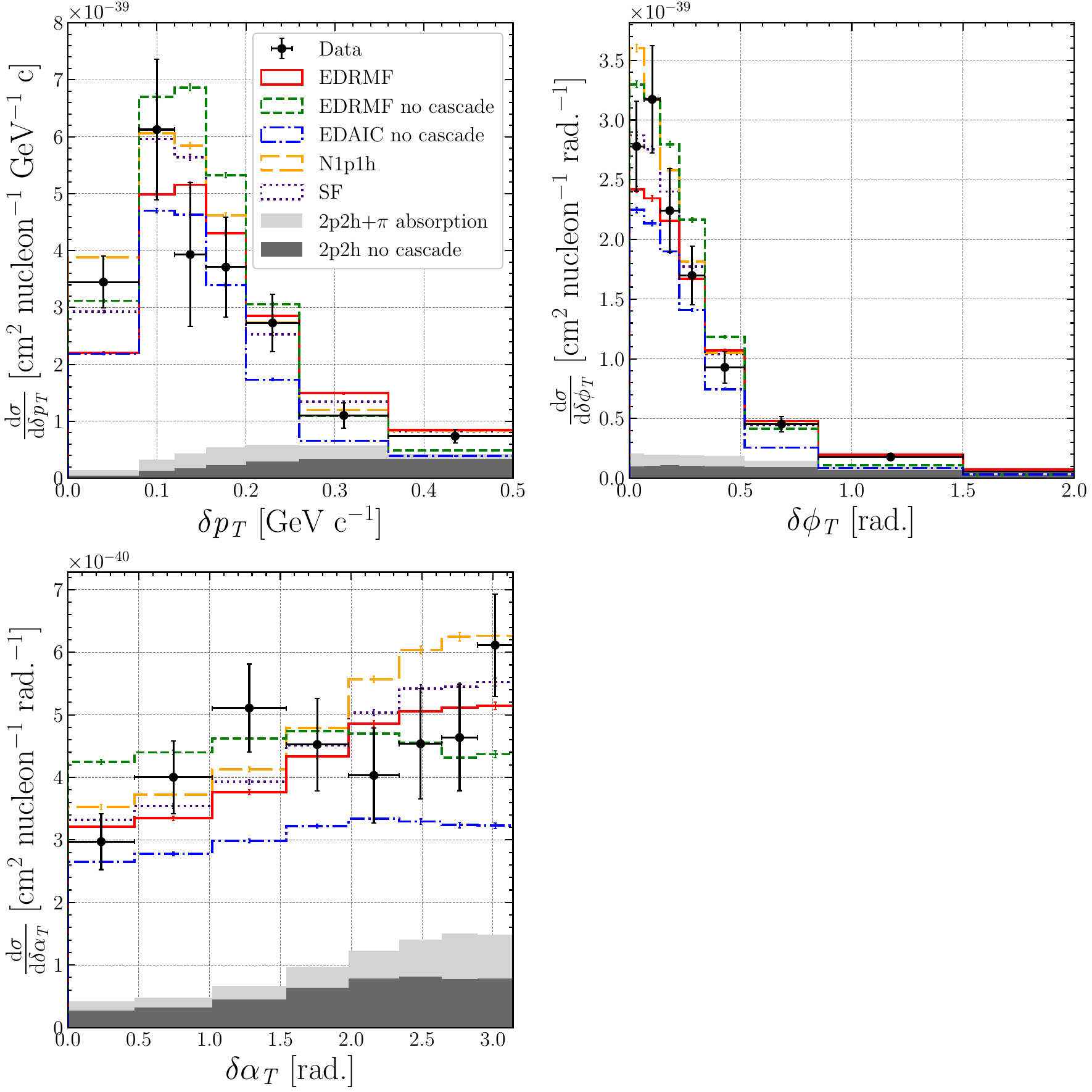}
    \caption{T2K $\nu_{\mu}$ CC$0\pi$N$p$ dataset on hydrocarbon. The TKI variables are defined in Eq.~\ref{eq:tki}. The histograms follow the same definition as Fig.~\ref{fig:t2k2018:1p_costheta}. Data taken from~\cite{T2K_2018_data}.}
    \label{fig:t2k2018:tki}
    \end{minipage}
\end{figure*}

\noindent
Fig.~\ref{fig:t2k2018:tki} shows the TKI variables for the leading proton and lepton, with the corresponding 
$\chi^2$ values listed in Table~\ref{table:T2K_TKI_chi2}. A shape-only comparison of the TKI variables is shown in Appendix~\ref{sec:app:F}. For 
$\delta p_T$, the EDRMF model with the cascade applied has the largest 
$\chi^2$ value compared to the existing models in NEUT. This discrepancy is likely due to the model underpredicting the data in the first bin at low $\delta p_T$, where CCQE interactions dominate, similar to the behaviour of the EDAIC model in this region. Conversely, the EDRMF model without the cascade applied shows the highest strength at the CCQE-dominated peak but underestimates the data at higher $\delta p_T$.\\

\noindent A similar trend is observed for $\delta \phi_T$, where the EDRMF model with the cascade applied again yields the largest $\chi^2$ value among the models in NEUT, primarily due to underestimating the data at low $\delta \phi_T$.\\

\noindent For the $\delta \alpha_T$ variable, the EDRMF model with the cascade applied shows a slight improvement in agreement with the data, as reflected in its $\chi^2$ value compared to other models. The cascade effect appears to reduce the predicted strength at lower $\delta \alpha_T$ while increasing it at higher $\delta \alpha_T$. However, none of the models successfully reproduce the structure observed in the data between 1 and 2 radians.

\begin{table} [H]
    \centering
    \begin{tabular}{  c |  c   c   c  }
    \hline \hline
    Model & $\delta p_{T}$ & $\delta \phi_{T}$ & $\delta \alpha_{T}$  \\ [0.5 ex]
    \hline \hline
    EDRMF cas & $42.1/8$ & $27.1/8$ & $17.6/8$\\
    EDRMF no cas & $60.4/8$ & $51.2/8$ & $27.5/8$\\
    N1p1h cas & $5.62/8$ & $12.6/8$ & $31.3/8$ \\
    SF cas & $11.3/8$ & $8.49/8$ & $19.4/8$\\
    EDAIC no cas & $24.6/8$ & $19.3/8$ & $18.8/8$ \\ [1ex]
    \hline
    \end{tabular}
    \caption{$\chi^{2}/N_{dof}$ values for each TKI variable for T2K. ``cas'' and ``no cas'' indicate where the cascade has and has not been applied respectively.
    } 
    \label{table:T2K_TKI_chi2}
\end{table}

\subsection{MINER$\nu$A data comparison}
\label{sec:Minerva}
\noindent CC0$\pi$Np dataset from Ref.~\cite{MinervaData:2018} with a correction applied in Ref.~\cite{minerva-data-set} is used to benchmark the model on hydrocarbon at higher neutrino flux peak energy. A comparison to the dataset before the correction was applied is shown in Appendix~\ref{sec:app:E}. The applied kinematic selections are given in Table~\ref{table:minerva_data_cuts}. \\

\begin{table} [H]
    \centering
    \begin{tabular}{ c | c }
    \hline \hline
    Kinematic Variable & CC0$\pi$Np \\ [0.5 ex]
    \hline \hline
    $p_{p}$ & 0.45 -- 1.2 GeV \\
    $\cos \theta_{p}$ & $>$ 0.3420 \\
    $p_{\mu}$ & 1.5 -- 10 GeV \\
    $\cos \theta_{\mu}$ & $>$ 0.9396 \\ [1ex]
    \hline
    \end{tabular}
    \caption{Kinematic selections applied in the MINER$\nu$A dataset. In the MINER$\nu$A analysis, any number of protons are selected within the kinematic selections, but only the highest momentum proton is used in the TKI and proton cross sections. A higher momentum proton may fall outside of the kinematic selections, however.}
    \label{table:minerva_data_cuts}
\end{table}

\subsubsection{Kinematic samples}
\begin{figure*}
    \begin{minipage}[h!]{\textwidth}
    \includegraphics[keepaspectratio, width=0.85\textwidth]{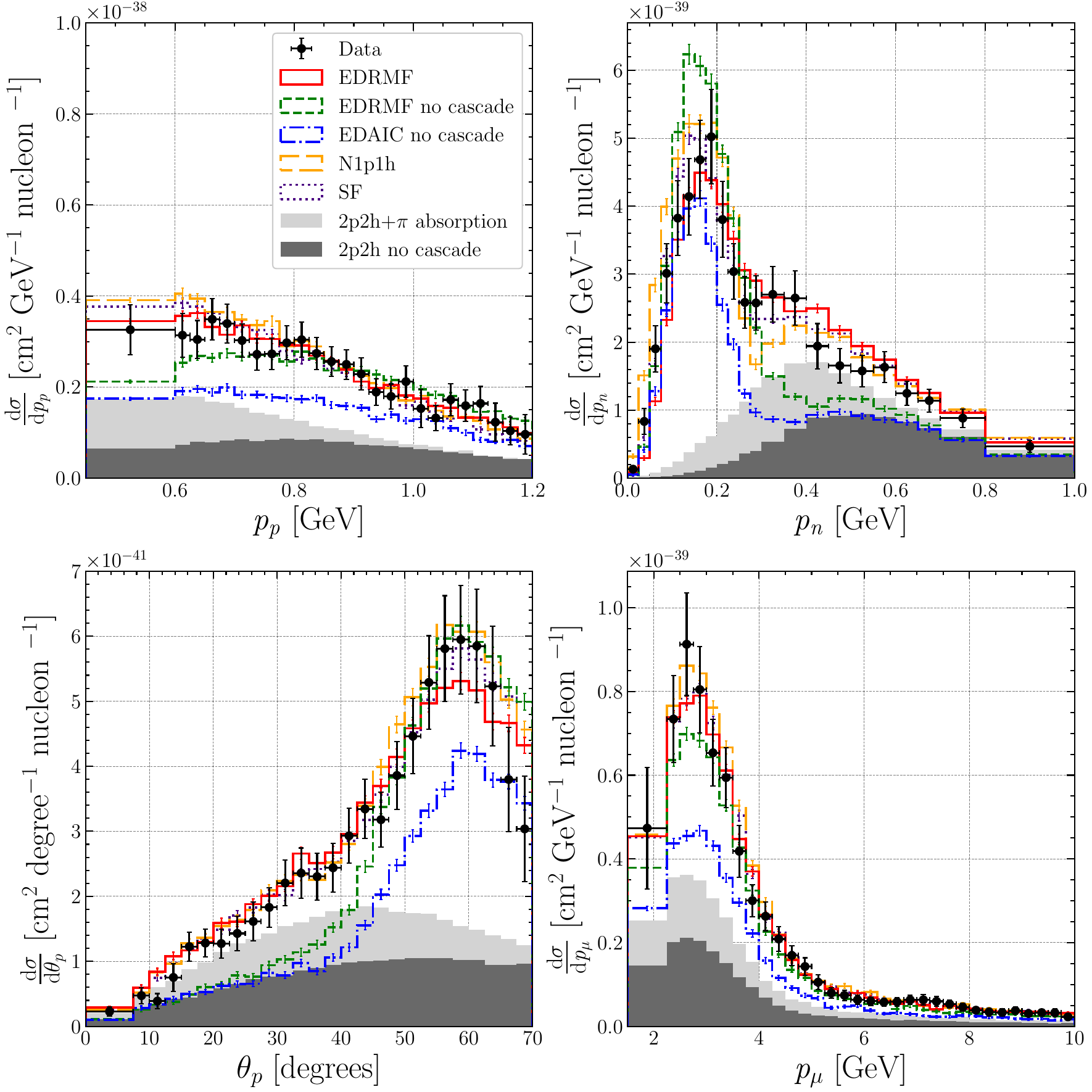}
    \caption{MINER$\nu$A $\nu_{\mu}$ CC$0\pi$N$p$ dataset on hydrocarbon. Differential cross sections for different kinematics are shown. Subscript $p$ indicated a scattered proton kinematic; a subscript $\mu$ indicates a scattered lepton kinematic and a subscript $n$ indicates a reconstructed initial neutron kinematic. The histograms follow the same definition as Fig.~\ref{fig:t2k2018:1p_costheta}. Data taken from~\cite{minerva-data-set}.}
    \label{fig:minerva:kinematics}
    \end{minipage}
\end{figure*}

\noindent Fig.~\ref{fig:minerva:kinematics} shows the differential cross-section distributions for the scattered muon momentum, leading scattered proton momentum, reconstructed initial neutron momentum, and leading proton scattering angle. The $\chi^{2}$ values are given in Table~\ref{table:minerva_kinematic_chi2}. A shape-only comparison of the kinematic variables is shown in Appendix~\ref{sec:app:F}. For the scattered muon momentum ($p_{\mu}$), the EDRMF model with cascade applied does not provide the best match to the data, as indicated by its higher $\chi^{2}$ value. We note that the EDRMF model without cascade has a better $\chi^{2}$ due to better agreement in the higher $p_{\mu}$ region.
In nearly all bins, the EDRMF model with the cascade applied shows the lowest cross section among all NEUT models with the cascade. \\

\noindent
For the scattered proton momentum ($p_{p}$), the EDRMF model with the cascade shows the best agreement with the data, as indicated by the $\chi^{2}$ value. The EDAIC model, despite not applying the cascade and exhibiting a lower normalisation compared to the data, achieves the second-best $\chi^{2}$ value. For proton scattering angle ($\theta_{p}$), the SF model shows the best agreement with the data based on the $\chi^{2}$ value. 
In the larger $\theta_{p}$ region, the EDRMF model with the cascade applied has the lowest strength. The EDAIC and the EDRMF model, without the cascade applied, exhibit similar predictions, indicating minimal CCQE contribution in this region. \\

\noindent
The reconstructed initial neutron momentum ($p_n$) is defined as~\cite{MinervaData:2018}: 
\begin{equation}\label{eq:pnreco-def}
    \begin{split}
        & p_{n} = \sqrt{\delta p_{L}^{2} + \delta p_{T}^{2}} \\
        & \delta p_{L} = \frac{1}{2}R - \frac{M_{B}^{2} + \delta p_{T}^{2}}{2R} \\
        & R = M_{A} + p_{L}^{\mu} + p_{L}^{p} - E^{\mu} - E^{p}
    .\end{split}
\end{equation}
\noindent Here, $L$, $B$, $\mu$ and $p$ indicate the longitudinal contribution, the residual nucleus, the muon and the proton respectively.
In $(e,e'p)$ experiments, the cross section in terms of missing momentum, which is analogous to $p_n$, is very sensitive to the description of the initial state and is closely related to the momentum distribution of nucleons in the nucleus. Hence, due to the different shapes of the momentum distributions of the shells $s$, $p$, $d$, etc., it allows one to identify the shell from which the scattered nucleon originated.
In our case, however, because of the average over the neutrino energy, all shells can contribute to a given $p_n$ value, and we do not see any shell structure in $d\sigma/dp_n$.\\

\noindent Despite this, $d\sigma/dp_n$ shows an interesting structure, with the CCQE clearly dominating below 0.2 GeV and the 2p2h and $\pi$-absorption providing the necessary strength above approximately 0.4 GeV, where there is very little CCQE.
In other words, $p_n$ is a powerful tool for discerning between CCQE and non-CCQE events. \\

\noindent In the CCQE dominated region, the N1p1h model clearly overestimates the data. The EDRMF with cascade, on the contrary, underestimates the data and provides a better description of the shape, having the best $\chi^2$ value among all models.

\begin{table} [H]
    \centering
    \begin{tabular}{  c  | c   c   c  c }
    \hline \hline
    Model & $p_{p}$ & $p_{n}$ & $\theta_{p}$ & $p_{\mu}$  \\ [0.5 ex]
    \hline \hline
    EDRMF cas & $26.2/25$ & $51.1/24$ & $54.1/26$ & $50.2/32$\\
    EDRMF no cas & $36.9/25$ & $112/24$ & $68.6/26$ & $43.9/32$\\
    N1p1h cas & $43.2/25$ & $105/24$ & $66.6/26$ & $41.6/32$\\
    SF cas & $45.4/25$ & $54.0/24$ & $40.9/26$ & $40.8/32$\\
    EDAIC no cas & $33.4/25$ & $52.4/24$ & $46.8/26$ & $37.1/32$\\ [1ex]
    \hline
    \end{tabular}
    \caption{$\chi^{2}/N_{dof}$ values for each kinematic variable variable for MINER$\nu$A. ``cas'' and ``no cas'' indicate where the cascade has and has not been applied respectively.
    }
    \label{table:minerva_kinematic_chi2}
\end{table}

\subsubsection{TKI samples}
\begin{figure*}
    \begin{minipage}[h!]{\textwidth}
    \includegraphics[keepaspectratio, width=0.85\textwidth]{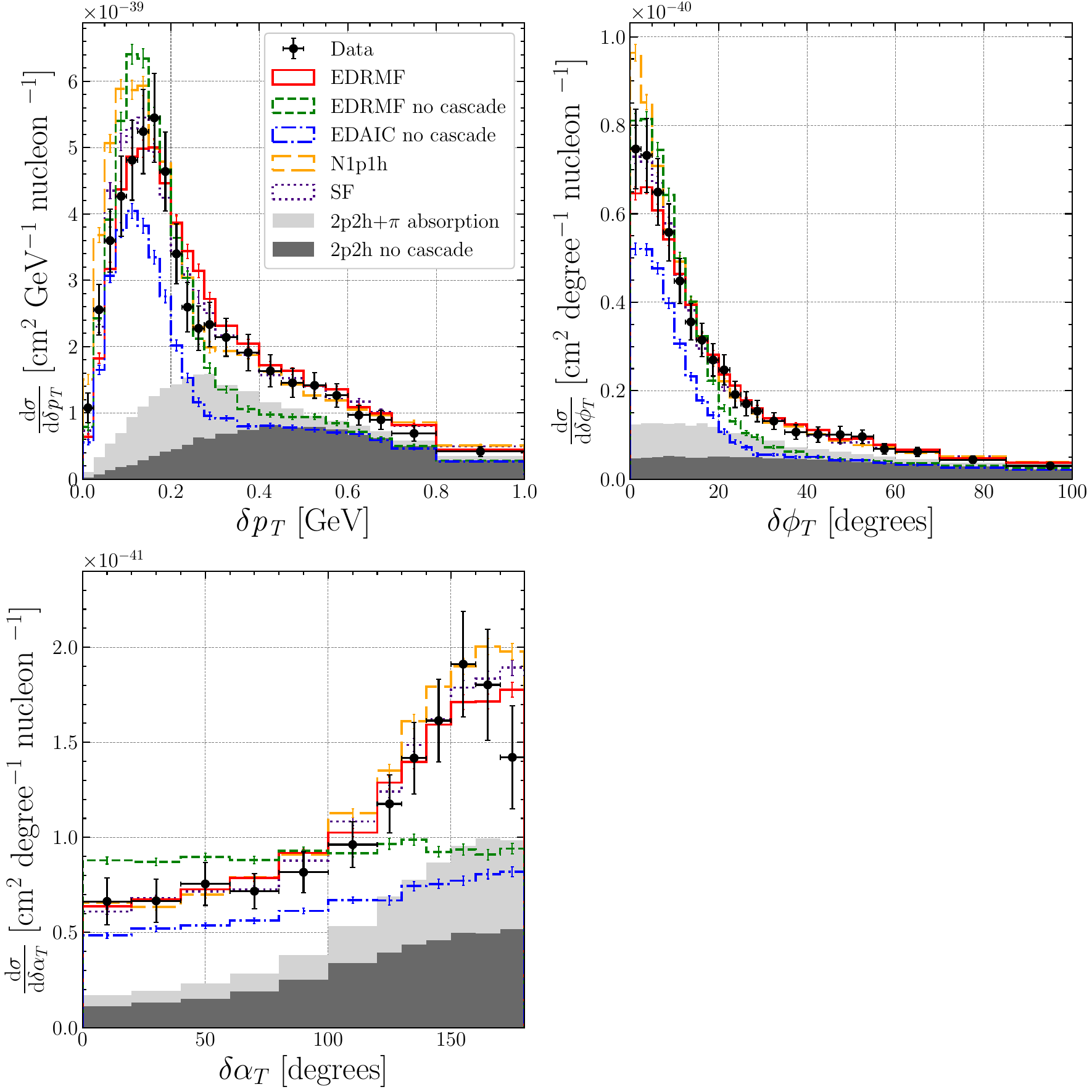}
    \caption{MINER$\nu$A $\nu_{\mu}$ CC$0\pi$N$p$ dataset on hydrocarbon. Differential cross sections for TKI variables defined in Eq.~\ref{eq:tki}. The histograms follow the same definition as Fig.~\ref{fig:t2k2018:1p_costheta}. Data taken from~\cite{minerva-data-set}.}
    \label{fig:minerva:TKI}
    \end{minipage}
\end{figure*}

\noindent Fig.~\ref{fig:minerva:TKI} shows the TKI cross sections. The $\chi^{2}$ values are given in Table~\ref{table:minerva_tki_chi2}. A shape-only comparison of the TKI variables is shown in Appendix~\ref{sec:app:F}.
$\delta p_{T}$ and $p_n$ are different representations of the missing momentum, so $d\sigma/d\delta p_{T}$ shows similar behaviour to $d\sigma/dp_n$ in Fig.~\ref{fig:minerva:kinematics}.
The EDRMF model with cascade applied has a significantly improved $\chi^{2}$ value, other models within NEUT overestimate the peak in a similar fashion to the EDRMF model without the cascade. However, the EDRMF model without cascade drastically underestimates the data after around 0.3~GeV$/c$.\\

\noindent For $\delta \phi_{T}$, the EDAIC model yields a better $\chi^{2}$ value, likely coming from the agreement in shape despite the normalisation being low to the eye. At low $\delta \phi_{T}$, when the angle between transverse $p_{p}$ and $-p_{\mu}$ is small (indicating an almost back-to-back proton and muon), the N1p1h model significantly overestimates the data. 
Beyond $20^{\circ}$, the 2p2h and $\pi$-absorption contributions from the cascade exceed those of the EDAIC model; however, this discrepancy is reduced at higher $\delta \phi_{T}$. \\

\noindent For $\delta \alpha_{T}$, the EDRMF model without the cascade exhibits a flat distribution. Without the cascade, the pion absorption contribution is omitted, leaving only pure CCQE and 2p2h events. The EDRMF model with the cascade applied yields a $\chi^2$ value slightly higher than those of the existing models in NEUT. Notably, all models incorporating the cascade overestimate the last bin at large $\delta \alpha_{T}$. 

\begin{table} [H]
    \centering
    \begin{tabular}{  c |  c   c   c  }
    \hline \hline
    Model & $\delta p_{T}$ & $\delta \phi_{T}$ & $\delta \alpha_{T}$  \\ [0.5 ex]
    \hline \hline
    EDRMF cas & $50.0/24$ & $54.0/23$ & $19.3/12$\\
    EDRMF no cas & $154/24$ & $102/23$ & $38.6/12$\\
    N1p1h cas & $182/24$ & $81.6/23$ & $18.5/12$ \\
    SF cas & $125/24$ & $71.9/23$ & $17.4/12$\\
    EDAIC no cas & $72.3/24$ & $46.7/23$ & $30.7/12$ \\ [1ex]
    \hline
    \end{tabular}
    \caption{$\chi^{2}/N_{dof}$ values for each TKI variable for MINER$\nu$A. ``cas'' and ``no cas'' indicate where the cascade has and has not been applied respectively.
    }
    \label{table:minerva_tki_chi2}
\end{table}

\subsection{MicroBooNE data comparison}
\label{sec:uBooNE}
\noindent CC$0\pi$1$p$ and CC$0\pi$N$p$ datasets~\cite{uboone-TKI:2023, uboone-TKI:2023:long, uboone-kinematics:2020} are used to benchmark the model on argon. The applied kinematic selections are given in Table~\ref{table:uboone_data_cuts}. \\

\begin{table} [H]
    \centering
    \begin{tabular}{ c | c c }
    \hline \hline
    Kinematic Variable & CC0$\pi$Np & TKI \\ [0.5 ex]
    \hline \hline
    $p_{p}$ & 0.3 -- 1.2 GeV & 0.3 -- 1 GeV \\
    $\cos \theta_{p}$ & $\dots$ & $\dots$ \\
    $p_{\mu}$ & $>$ 0.1 GeV & 0.1 -- 1.2 GeV \\
    $\cos \theta_{\mu}$ & $\dots$ & $\dots$ \\ [1ex]
    \hline
    \end{tabular}
    \caption{Kinematic selections applied in the MicroBooNE dataset. Here ``$\dots$'' indicate that there is no selection applied. In the MicroBooNE analysis, only one proton is selected in the kinematic region for the TKI cross sections. This selected proton may not be the highest momentum proton in the event as this may be outside the kinematic region considered. Events with charged pions may pass selection criteria if their momentum is below 70 MeV. In the N$p$ analysis, any number of protons can be selected, but only the highest momentum proton is used to calculate the proton cross sections. There are no pions or other mesons in the final state.}
    \label{table:uboone_data_cuts}
\end{table}

\subsubsection{Kinematic samples}
\begin{figure*}
    \begin{minipage}[h!]{\textwidth}
    \includegraphics[keepaspectratio, width=0.85\textwidth]{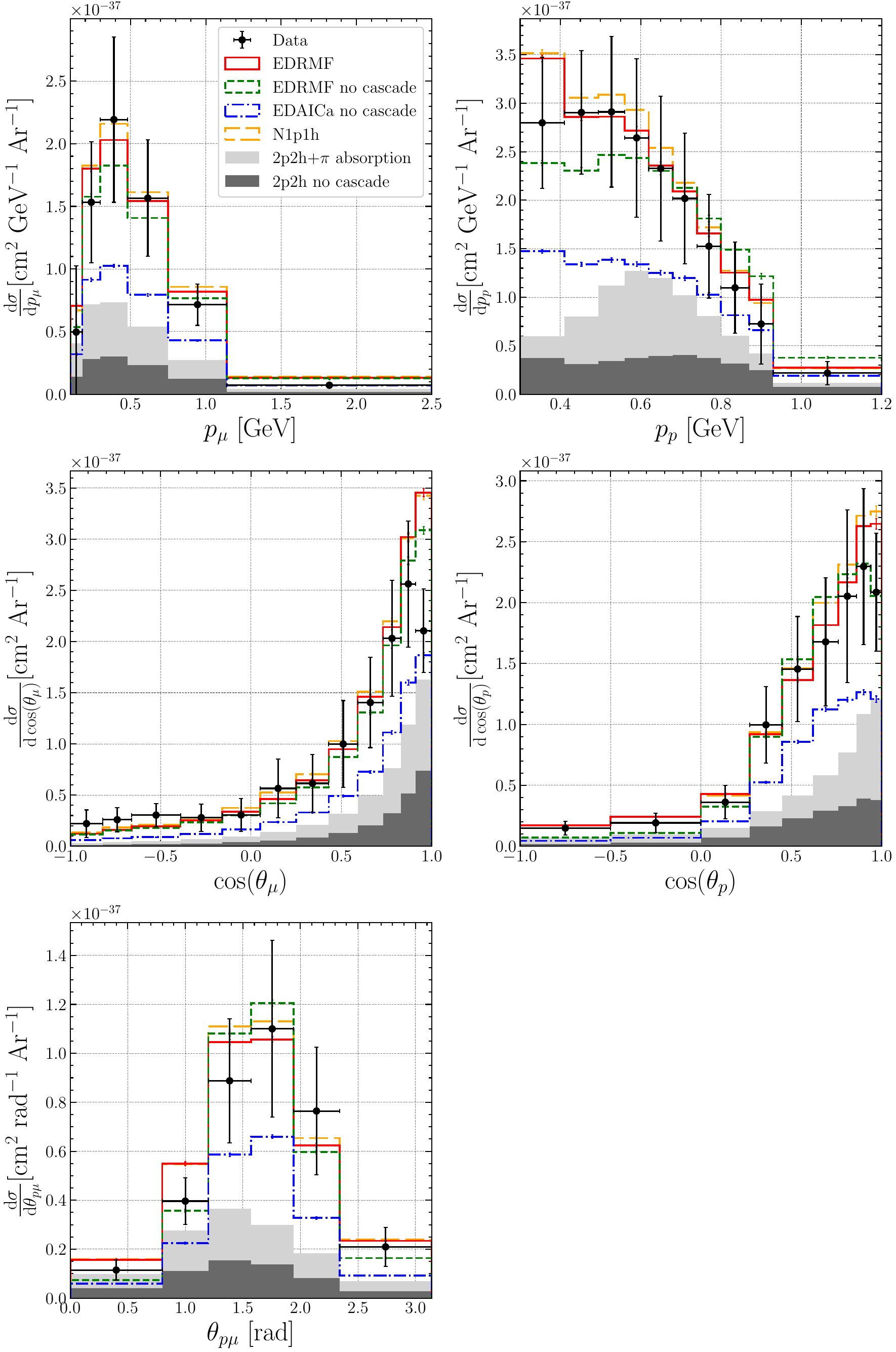}
    \caption{MicroBooNE $\nu_{\mu}$ CC$0\pi$N$p$ dataset on argon. Differential cross sections for different kinematics are shown. Subscript $p$ indicated a scattered proton kinematic; a subscript $\mu$ indicates a scattered muon kinematic and the subscript $p\mu$ indicates the opening angle between the scattered proton and muon. The histograms follow the same definition as Fig.~\ref{fig:t2k2018:1p_costheta} except the SF model is not included. Data taken from~\cite{uboone-kinematics:2020, uboone-TKI:2023:long}.}
    \label{fig:uboone:kinematics}
    \end{minipage}
\end{figure*}

\noindent Fig.~\ref{fig:uboone:kinematics} presents the differential cross section for the following observables: the scattered muon momentum ($p_{\mu}$), leading scattered proton momentum ($p_{p}$), muon scattering angle ($\theta_{\mu}$), leading proton scattering angle ($\theta_{p}$), and the opening angle between the scattered proton and muon ($\theta_{p\mu}$). The corresponding $\chi^2$ values are summarised in Table~\ref{table:uboone_kinematic_chi2}. A shape-only comparison of the kinematic variables is shown in Appendix~\ref{sec:app:F}. \\

\noindent For the scattered muon momentum, the EDAICa model achieves the lowest $\chi^2$, indicating the best agreement with the data. Although the EDAICa model appears to have a lower normalisation by eye, high correlations in data can result in a relatively good $\chi^2$ if the overall shape is well reproduced. In contrast, the other models have larger $\chi^2$ values, suggesting poorer agreement with the data. Visual inspection shows that the EDRMF model with cascade and the N1p1h model both provide a reasonable match to the data, although there is a slight difference in the peak position. This level of agreement was less pronounced in comparisons with MINER$\nu$A and T2K data.\\

\noindent For the scattered proton momentum, the EDRMF model with the cascade applied and the N1p1h model have the best $\chi^{2}$ values.
Both of these models also show an increase in strength in the first bin. This is not seen in models that do not have the cascade applied, indicating that the cascade may be increasing the strength too much in this first bin. The EDAICa model without the cascade applied has the lowest strength as expected. \\

\noindent For $\cos(\theta_{\mu})$, the EDAICa model achieves the best $\chi^{2}$ value. In the last bin, the N1p1h and EDRMF models with the cascade overestimate the data and this could be a reason for the increase $\chi^{2}$ value for those models. In the last bin, all models fail to replicate the drop in shape. It is however not possible to conclude whether this excess is due to the CCQE or to the non-CCQE contributions, which are quite large in the forwardest bin.\\

\noindent For $\cos(\theta_{p})$, the EDRMF model with the cascade applied is the best fit to data given the $\chi^{2}$ and the EDAICa model without the cascade is the next best. It is also interesting to note that the EDRMF and EDAICa model without the cascade applied replicates the drop in the last bin, whilst models with the cascade applied does not. Similarly to $\cos(\theta_{\mu})$, the last bin is overestimated by N1p1h and EDRMF with the cascade applied. This indicates that the NEUT cascade is increasing the strength in the last $\cos(\theta_\mu)$ and $\cos(\theta_p)$ bins and could possibly be due to the non-CCQE contributions.\\

\noindent For the muon-proton opening angle, the EDAICa model without the cascade is the best fit to the data given the $\chi^{2}$. Again, the strong correlations in the data may contribute to this outcome. 

\begin{table} [H]
    \centering
    \begin{tabular}{  c |  c   c   c  c  c }
    \hline \hline
    Model & $p_{\mu}$ & $p_{p}$ & $\cos(\theta_{\mu})$ & $\cos(\theta_{p})$ & $\theta_{p \mu}$ \\ [0.5 ex]
    \hline \hline
    EDRMF cas & $26.7/6$ & $7.74/10$ & $21.9/12$ & $7.57/9$ & $10.6/6$\\
    EDRMF no cas & $28.2/6$ & $19.4/10$ & $14.8/12$ & $14.4/9$ & $13.4/6$\\
    N1p1h cas & $28.6/6$ & $4.09/10$ & $21.1/12$ & $9.20/9$ & $9.48/6$\\
    EDAICa no cas & $10.7/6$ & $17.4/10$ & $8.34/12$ & $8.17/9$ & $5.73/6$\\ [1ex]
    \hline
    \end{tabular}
    \caption{$\chi^{2}/N_{dof}$ values for each kinematic variable variable for MicroBooNE. ``cas'' and ``no cas'' indicate where the cascade has and has not been applied respectively.
    }
    \label{table:uboone_kinematic_chi2}
\end{table}

\subsubsection{TKI samples}
\begin{figure*}
    \begin{minipage}[h!]{\textwidth}
    \includegraphics[keepaspectratio, width=0.85\textwidth]{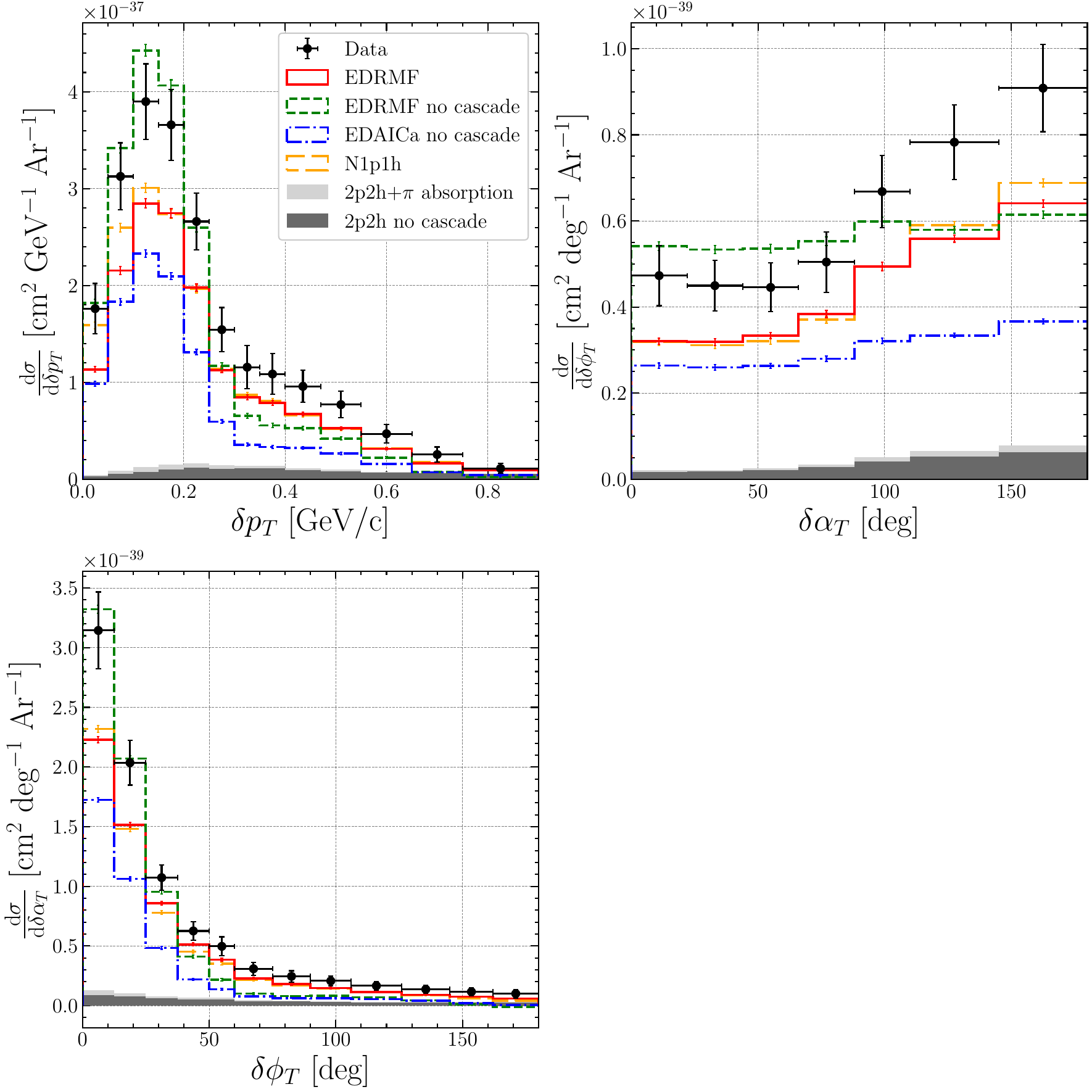}
    \caption{Differential cross sections for TKI variables defined in Eq.~\ref{eq:tki} for MicroBooNE. The histograms follow the same definition as Fig.~\ref{fig:t2k2018:1p_costheta} except the SF model is not included. Correlations in data lead to a visual normalisation difference between the MC and measurement. Data taken from~\cite{uboone-kinematics:2020}.}
    \label{fig:uboone:TKI}
    \end{minipage}
\end{figure*}

\noindent Fig.~\ref{fig:uboone:TKI} shows the TKI cross sections. The corresponding $\chi^{2}$ values are given in Table~~\ref{table:uboone_tki_chi2}. A shape-only comparison of the TKI variables is shown in Appendix~\ref{sec:app:F}.  
For $\delta p_{T}$, the EDRMF model with the cascade applied achieves the lowest $\chi^2$ value, while the N1p1h model also shows good agreement with the data, as indicated by a reduced $\chi^2$ slightly above 1. Both the EDRMF and EDAICa models without the cascade show a high $\chi^{2}$ value. It can be seen  that all models except the EDRMF without the cascade applied underestimate the data in all bins. The EDRMF model, without the cascade applied, is consistent with data in the peak of the distribution and begins to underestimate the data in the sixth bin. \\

\noindent For $\delta \alpha_{T}$, both the N1p1h model and the EDRMF model with the cascade show good agreement with the data based on their $\chi^2$ values.
For $\delta \phi_{T}$, the $\chi^{2}$ values indicate that the EDRMF model with the cascade provides the best agreement with the data, while the N1p1h model achieves the second-best agreement. The EDRMF model without the cascade applied is more consistent with the data in the low-angle region compared to other models, all of which underestimate the data.

\begin{table} [H]
    \centering
    \begin{tabular}{  c | c   c   c  }
    \hline \hline
    Model & $\delta p_{T}$ & $\delta \phi_{T}$ & $\delta \alpha_{T}$  \\ [0.5 ex]
    \hline \hline
    EDRMF cas & $11.0/13$ & $12.0/12$ & $8.34/7$\\
    EDRMF no cas & $26.7/13$ & $32.1/12$ & $50.3/7$\\
    N1p1h cas & $13.3/13$ & $13.5/12$ & $7.35/7$ \\
    EDAICa no cas & $26.1/13$ & $38.3/12$ & $31.6/7$ \\ [1ex]
    \hline
    \end{tabular}
    \caption{$\chi^{2}/N_{dof}$ values for each TKI variable for MicroBooNE. ``cas'' and ``no cas'' indicate where the cascade has and has not been applied respectively.
    }
    \label{table:uboone_tki_chi2}
\end{table}

\section{Conclusion}
\label{sec:conclusion}
\noindent A model based on the use of relativistic mean field theory and relativistic distorted waves to describe the initial and scattered state has been implemented into the NEUT neutrino event generator for the CCQE scattering channel. The model includes different nuclear potentials and accounts, in a relativistic and quantum mechanical framework, for binding energies, the Pauli exclusion principle, Fermi motion and the mean-field distortion of the nucleon wave functions. It incorporates nuclear effects beyond the independent-particle shell model by using a semi-phenomenological spectral function. It provides a reasonable description of inclusive and exclusive electron scattering data in the kinematic region dominated by the quasi-elastic peak.\\ 

\noindent The model was implemented by use of precomputed hadronic tensor tables. The model has been benchmarked using T2K, MINER$\nu$A CC0$\pi$Np and MicroBooNE cross-section data. In the case of T2K, the EDRMF model including inelastic FSI from NEUT's cascade model has the lowest strength in bins that are not dominated by 2p2h and $\pi$-absorption contributions. \\

\noindent For the case of MINER$\nu$A, the neutrino flux has a higher average energy so contributions from 2p2h and $\pi$ absorption are more prominent. There is a significant improvement in $\chi^{2}$ values for the EDRMF model with the cascade applied when compared to other models within NEUT, the largest of which is seen in reconstructed neutron momenta. It is evident that the initial-state nuclear modelling using RMF theory with a phenomenological spectral function and the distortion of the outgoing nucleon provides an improved description of the reconstructed initial nucleon kinematics. The TKI distributions, specifically the $\delta p_{T}$ distribution, also show an improvement in $\chi^{2}$ for the EDRMF model with the cascade applied. \\

\noindent For the case of MicroBooNE, the kinematic cross sections show that the EDAICa model without the cascade applied provides an improved $\chi^{2}$. This is initially surprising given the visually low normalisation. However, the covariance matrices accompanying the data show correlations which can result in an improved $\chi^{2}$ for visually low normalisations if the shape is well captured. We observe that all models with the cascade applied predict significantly higher cross sections at forward muon and proton angles compared to both T2K and MicroBooNE data, as shown in Figures \ref{fig:t2k2018:1p_costhetap_slices} and \ref{fig:uboone:kinematics}. It is important to note that in these regions the contribution from $\pi$-absorption is larger. The TKI measurements show an improved $\chi^{2}$ for the $\delta p_{T}$ and $\delta \phi_{T}$ distributions. For $\delta \alpha_{T}$, the N1p1h model has the best $\chi^{2}$. By eye, the N1p1h and the EDRMF model with the cascade applied agree until the last two bins; this region could result in the improved $\chi^{2}$ for the N1p1h model. \\

\noindent The observed lack of strength in the region of phase space where the contributions of 2p2h and $\pi$-absorption are small suggests that additional unaccounted-for effects that increase the CCQE cross section could be incorporated into this model.
In particular, we note that all the RDWIA model predictions presented here have been calculated using a dipole parametrisation for the axial-vector form factor of the nucleon with mass parameter $M_A^{QE} = 1.05$~GeV. 
However, recent findings from lattice QCD calculations~\cite{Meyer22} and the MINER$\nu$A analysis based on antineutrino-proton scattering data~\cite{Cai23} suggest that the dipole parameterisation fails to accurately describe the $Q^2$ behaviour of the form factor. These studies indicate that the form factor at $Q^2 > 0$ is significantly larger than the dipole prediction with $M_A^{QE} \approx 1$~GeV, leading to a substantially higher CCQE cross section.
An additional potential source of increased CCQE cross section arises from the interference between one- and two-body currents. This has been explored extensively in electron scattering studies~\cite{Dekker92, VanderSluys95, Carlson02, Lovato16, Franco-Munoz23, Franco-Munoz25, Lovato23, Andreoli24} and in weak current interactions~\cite{Lovato14, Lovato23, Franco-Munoz25b}, which suggests that such interference effects could play a significant role.\\

\noindent The implemented model can be extended in future work to include short-ranged correlations by explicitly modelling the ejection of two nucleons. Additionally, the precomputed hadronic tensor tables can be reparameterised to separate the individual contributions to the hadronic tensor: vector-vector, vector-axial, and axial-axial. 
This separation gives more freedom for modelling and would allow, e.g., to change the axial-vector form factor without the need to recompute the hadronic tensor tables.\\

 \noindent The use of precomputed tables, however, still considerably restricts the evaluation of systematic uncertainties. For example, many tables with different RMF model parameters would be required to estimate the uncertainty in the initial state. A different approach would be to precompute and tabulate the final-state spinor. This would enable direct changes to the RMF model parameters without the need for recomputing the spinor table. A drawback of this approach is that there is a three-dimensional integral that must be solved per event to construct the hadronic current. At this moment, the precomputed hadronic tensor table is the more efficient approach.  \\

\noindent
In addition, the nuclear model can be applied to describe single pion production (SPP) in lepton scattering on nuclear targets. Accurately modelling such interactions is critical for all neutrino experiments, particularly for future experiments like DUNE, where neutrino interactions are dominated by these processes. For this extension, the CCQE transition operator in Eq.~\ref{CCQE_TO} would need to be replaced with SPP transition operators as discussed in Ref.~\cite{pi-Gonzalez-Jimenez:2017fea}. These operators, already developed in Refs.~\cite{hybrid-Gonzalez-Jimenez:2016qqq, Kabirnezhad:2017jmf, Kabirnezhad:2020wtp, Kabirnezhad:2022znc, Kabirnezhad:2024cor} for a broad kinematic region, meet the requirements of DUNE's wide neutrino beam spectrum. \\

\section{Acknowledgments}
\label{sec:Acknowledgments}
\noindent We acknowledge the support by the Royal Commission for the Exhibition of 1851, the STFC and UKRI, UK. We also thank the T2K Neutrino Interactions Working Group for their valuable discussions and feedback throughout the implementation of this model in NEUT, with special appreciation to Dr. Luke Pickering for his invaluable expertise on NEUT. R.G.-J. was supported by projects PID2021-127098NA-I00 and RYC2022-035203-I funded by MICIU/AEI/10.13039/501100011033, ``ERDF a way of making Europe'' and FSE+.

% \bibliography{bibliography.bib}

\clearpage

% \appendix
\begin{appendices}

\section{}
\label{sec:app:A}

\subsection{Carbon missing energy density}
\begin{figure}[H]
    \centering
    \includegraphics[keepaspectratio, width=0.5\textwidth]{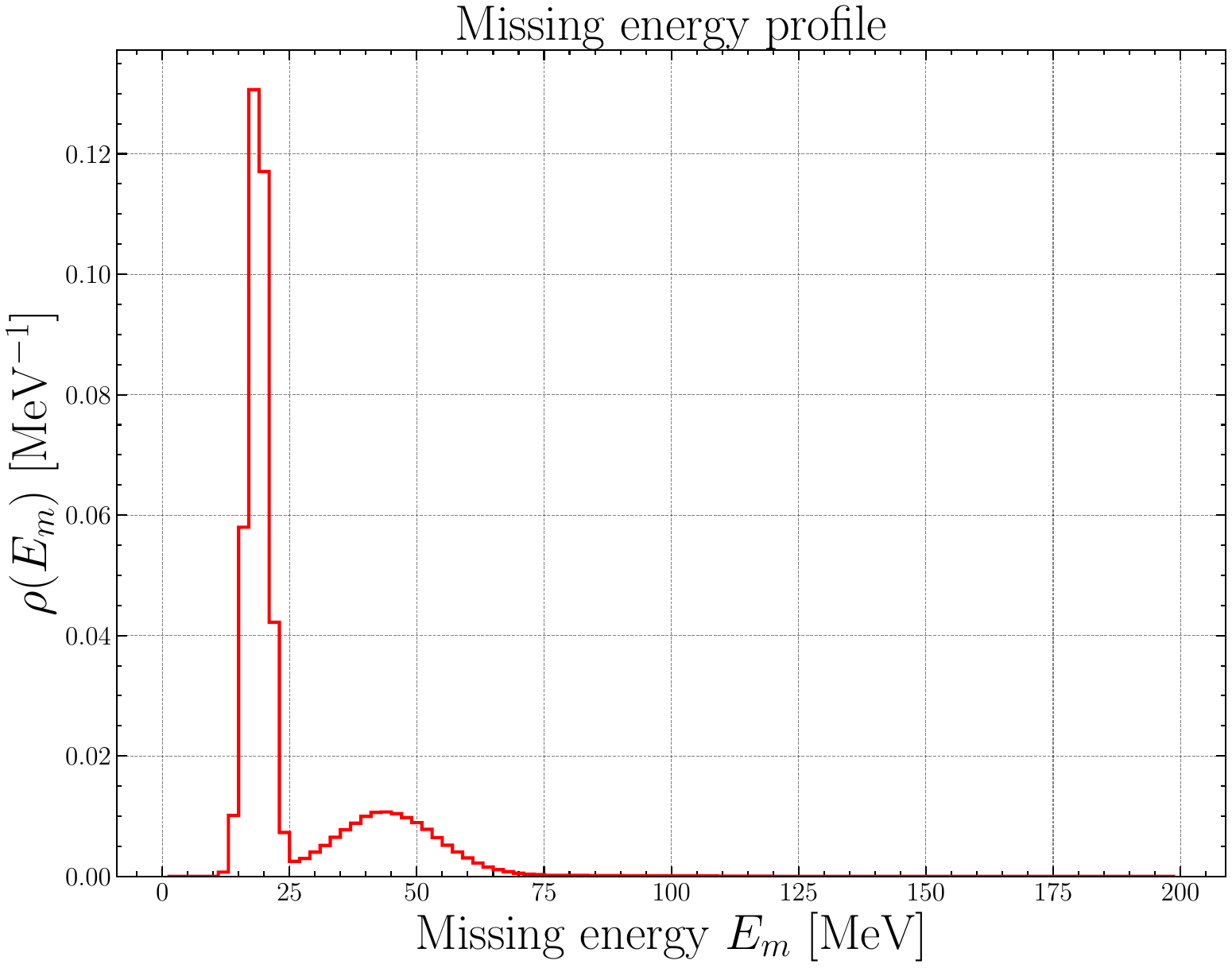}
    \caption{Missing energy profile for $^{12}$C. The two $1p_{3/2}$ and $1s_{1/2}$ shells are clearly seen as the large, narrow peak and broad lower peak respectively.}
    \label{fig:emiss_prof_c12}
\end{figure}

\noindent Nucleons in Carbon occupy the $1s_{1/2}$ and $1p_{3/2}$ shells. For these shells, the missing energy density is given as 

\begin{equation}
\rho_{\kappa}(E_{m}) = \frac{N_{\kappa}}{\sqrt{2\pi}\sigma^{\kappa}} e^{-\frac{(E_{m} - E_{m}^{\kappa})^{2}}{2(\sigma^{\kappa})^{2}}},
\end{equation}

\noindent where $N_{\kappa}$ is the occupancy of the shell $\kappa$. The central values and widths of these Gaussian functions are given below in Table~\ref{table:proton_c12_Em_prof} and Table~\ref{table:neutron_c12_Em_prof}. The normalised missing energy distribution for protons is shown in Fig.~\ref{fig:emiss_prof_c12}.  
The central values of the Gaussian functions correspond to the eigenvalues predicted by the RMF model with the NLSH parameter set~\cite{NLSH}. The width and occupations are from the Rome spectral function for carbon~\cite{Benhar94,Benhar05,Benhar08} following the procedure described in~\cite{Franco-Patino22}.
\\

\begin{table} [H]
    \centering
    \begin{tabular}{  c c c c }
    \hline \hline
    Shell & Occupancy & $E_{m}^{\kappa}$[MeV] & $\sigma^{\kappa}$[MeV] \\ [0.5 ex]
    \hline \hline
    $1s_{1/2}$ & 1.9/2 & 40.5 & 10.0 \\
    $1p_{3/2}$ & 3.3/4 & 14.46 & 2.0   \\ [1ex]
    \hline
    \end{tabular}
    \caption{Central values and widths of Gaussian functions used to model the proton shells in $^{12}$C.}
    \label{table:proton_c12_Em_prof}
\end{table}

\begin{table} [H]
    \centering
    \begin{tabular}{  c c c c }
    \hline \hline
    Shell & Occupancy & $E_{m}^{\kappa}$[MeV] & $\sigma^{\kappa}$[MeV] \\ [0.5 ex]
    \hline \hline
    $1s_{1/2}$ & 1.9/2 & 44.37 & 10.0 \\
    $1p_{3/2}$ & 3.3/4 & 17.81 & 2.0   \\ [1ex]
    \hline
    \end{tabular}
    \caption{Central values and widths of Gaussian functions used to model the neutron shells in $^{12}$C.}
    \label{table:neutron_c12_Em_prof}
\end{table} 

\subsection{Oxygen missing energy density}
\begin{figure}[H]
    \centering
    \includegraphics[keepaspectratio, width=0.5\textwidth]{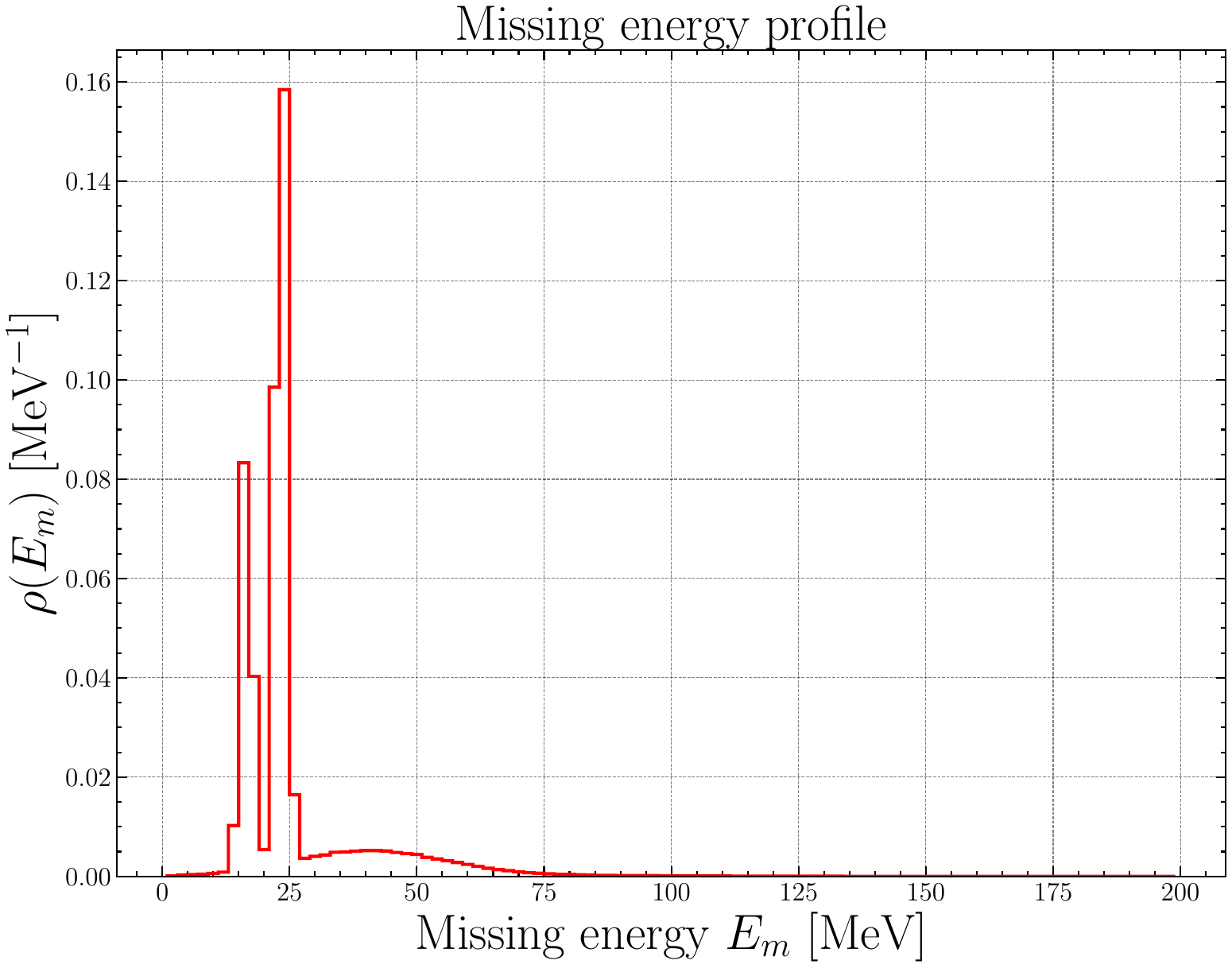}
    \caption{Missing energy profile for $^{16}$O. The two $1p_{3/2}$ and $1p_{1/2}$ shells are clearly seen as the large, narrow peaks and the $1s_{1/2}$ is the broad lower peak respectively.}
    \label{fig:emiss_prof_o16}
\end{figure}

\noindent Nucleons in Oxygen occupy the $1s_{1/2}$, $1p_{3/2}$ and $1p_{1/2}$ shells. The central values and widths of the Gaussian functions are given below in Table~\ref{table:proton_o16_Em_prof} and Table~\ref{table:neutron_o16_Em_prof}. The normalised missing energy density is shown in Fig.~\ref{fig:emiss_prof_o16}. The central values of the Gaussian functions correspond to the eigenvalues predicted by the RMF model with the set of NLSH parameters~\cite{NLSH}.
The width and occupations are from the Rome
spectral function for oxygen~\cite{Benhar94,Benhar05,Benhar08}
following the procedure described in~\cite{Gonzalez-Jimenez22}.
\\

\begin{table} [H]
    \centering
    \begin{tabular}{  c c c c }
    \hline \hline
    Shell & Occupancy & $E_{m}^{\kappa}$[MeV] & $\sigma^{\kappa}$[MeV] \\ [0.5 ex]
    \hline \hline
    $1s_{1/2}$ & 1.62/2 & 37.70 & 15.0 \\
    $1p_{3/2}$ & 3.47/4 & 18.27 & 1.0 \\
    $1p_{1/2}$ & 1.51/2 & 11.49 & 1.0   \\ [1ex]
    \hline
    \end{tabular}
    \caption{Central values and widths of Gaussian functions used to model the proton shells in $^{16}$O.}
    \label{table:proton_o16_Em_prof}
\end{table}

\begin{table} [H]
    \centering
    \begin{tabular}{  c c c c }
    \hline \hline
    Shell & Occupancy & $E_{m}^{\kappa}$[MeV] & $\sigma^{\kappa}$[MeV] \\ [0.5 ex]
    \hline \hline
    $1s_{1/2}$ & 1.62/2 & 42.24 & 15.0 \\
    $1p_{3/2}$ & 3.47/4 & 22.36 & 1.0 \\
    $1p_{1/2}$ & 1.51/2 & 15.49 & 1.0   \\ [1ex]
    \hline
    \end{tabular}
    \caption{Central values and widths of Gaussian functions used to model the neutron shells in $^{16}$O.}
    \label{table:neutron_o16_Em_prof}
\end{table}

\subsection{Argon missing energy density}
\begin{figure}[H]
    \centering
    \includegraphics[keepaspectratio, width=0.5\textwidth]{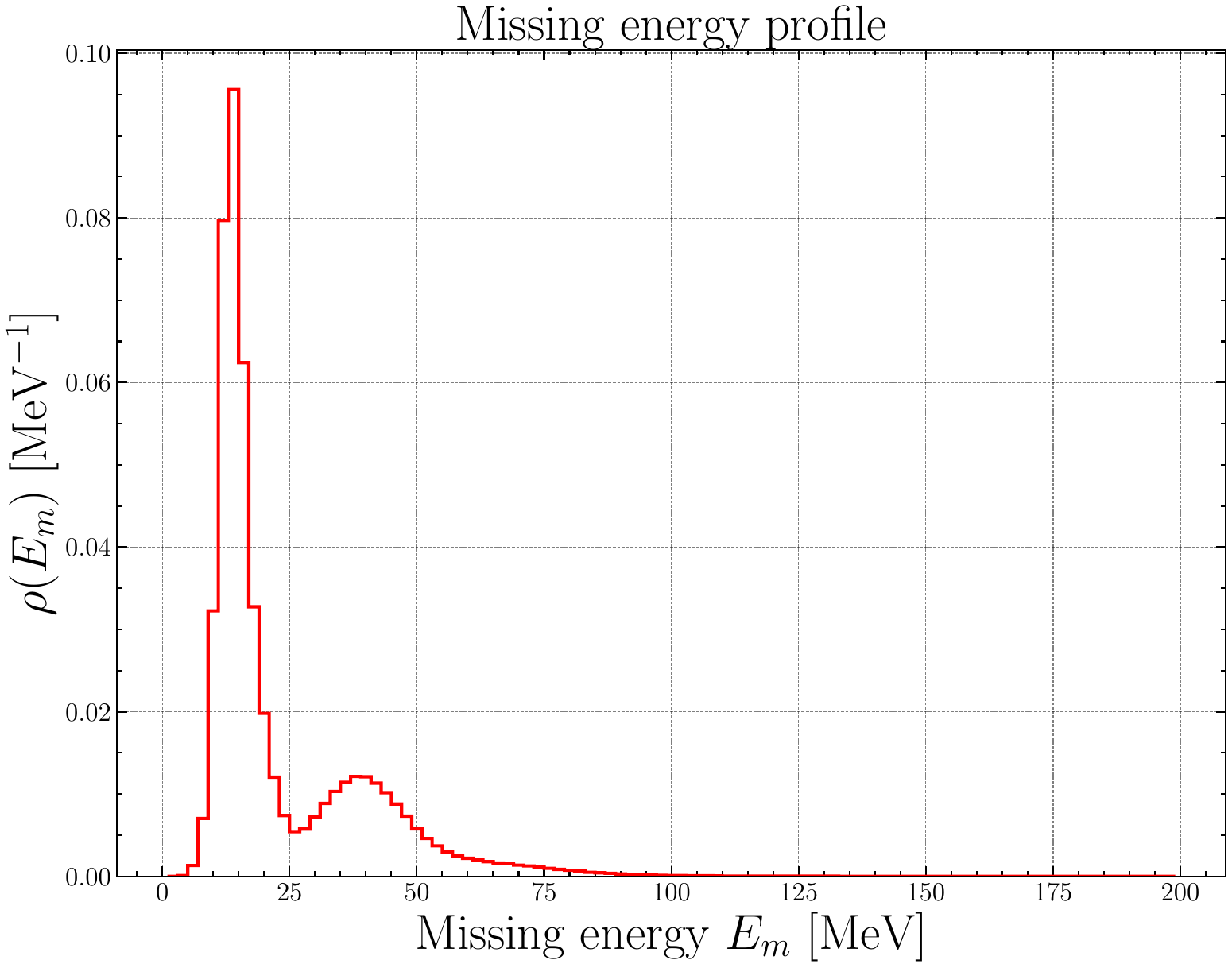}
    \caption{Missing energy profile for $^{40}$Ar.}
    \label{fig:emiss_prof_Ar40}
\end{figure}

\noindent Nucleons in Argon occupy the $1s_{1/2}$, $1p_{3/2}$, $1p_{1/2}$, $1d_{5/2}$, $1d_{3/2}$ and $2s_{1/2}$ shells. The central values and widths of the Gaussian functions are given below in Table~\ref{table:proton_Ar40_Em_prof} and Table~\ref{table:neutron_Ar40_Em_prof}. The normalised missing energy density is shown in Fig.~\ref{fig:emiss_prof_Ar40}. The central values of the Gaussian functions correspond to the eigenvalues predicted by the RMF model with the set of NLSH parameters~\cite{NLSH}. 
The occupancies are taken from~\cite{Butkevich12}. 
For the widths, we use thinner distributions for the external shells and wider distributions for the internal ones. 
We point out that the structure of $^{40}$Ar is not as constrained as $^{12}$C and $^{16}$O; therefore, one should consider large uncertainties attached to this missing energy profile.
\\
\begin{table} [H]
    \centering
    \begin{tabular}{  c c c c }
    \hline \hline
    Shell & Occupancy & $E_{m}^{\kappa}$[MeV] & $\sigma^{\kappa}$[MeV] \\ [0.5 ex]
    \hline \hline
    $1s_{1/2}$ & 2.0/2 & 47.42 & 8.0 \\
    $1p_{3/2}$ & 3.8/4 & 32.31 & 8.0 \\
    $1p_{1/2}$ & 1.9/2 & 27.72 & 8.0 \\
    $1d_{5/2}$ & 4.8/6 & 17.14 & 4.0 \\
    $1d_{3/2}$ & 1.7/4 & 10.02 & 2.0 \\
    $2s_{1/2}$ & 1.7/2 & 9.89 & 2.0 \\[1ex]
    \hline
    \end{tabular}
    \caption{Central values and widths of Gaussian functions used to model the proton shells in $^{40}$Ar.}
    \label{table:proton_Ar40_Em_prof}
\end{table}

\begin{table} [H]
    \centering
    \begin{tabular}{  c c c c }
    \hline \hline
    Shell & Occupancy & $E_{m}^{\kappa}$[MeV] & $\sigma^{\kappa}$[MeV] \\ [0.5 ex]
    \hline \hline
    $1s_{1/2}$ & 2.0/2 & 54.56 & 15.0 \\
    $1p_{3/2}$ & 3.8/4 & 38.74 & 8.0 \\
    $1p_{1/2}$ & 1.9/2 & 34.20 & 8.0 \\
    $1d_{5/2}$ & 4.8/6 & 23.09 & 4.0 \\
    $1d_{3/2}$ & 3.3/4 & 15.95 & 2.0 \\
    $2s_{1/2}$ & 1.7/2 & 16.11 & 2.0 \\
    $1f_{7/2}$ & 1.6/8 & 8.27 & 2.0 \\[1ex]
    \hline
    \end{tabular}
    \caption{Central values and widths of Gaussian functions used to model the neutron shells in $^{40}$Ar.}
    \label{table:neutron_Ar40_Em_prof}
\end{table}

\noindent \noindent The normalised missing energy profile is shown in Fig.~\ref{fig:emiss_prof_Ar40}

\subsection{Background shell modelling}
\noindent The additional $1s_{1/2}$ shell is modelled differently for different $E_{m}$ regions. \\

\underline{$E_{m} < 26$ MeV}:\\
\begin{equation}
\rho(E_{m}) = 0.
\end{equation}

\underline{$26 < E_{m} < 100$ MeV}:\\
\begin{equation}
\rho(E_{m}) = \frac{ae^{-100b}}{e^{-(E_{m} - c)/w} + 1}.
\end{equation}

\underline{$E_{m} > 100$ MeV}:\\
\begin{equation}
\rho(E_{m}) = ae^{-bE_{m}}.
\end{equation}

\noindent $26$ MeV is an estimated value for the two-nucleon knockout threshold. This value is different for each nuclei, and for each isospin combination of the nucleon pair. In this work, for simplicity, we used the same value for the three studied nuclei and neglected isospin dependence. \\

\noindent For oxygen, the parameters are described in Table~\ref{table:background_shell_params}. The same functional form is used for the three studied nuclei but it is re-scaled to give the correct number of nucleons (Table~\ref{table:background_shell_occup}). \\

\begin{table} [H]
    \centering
    \begin{tabular}{  c c  }
    \hline \hline
    Parameter & Value  \\ [0.5 ex]
    \hline \hline
    $a$ & 0.03113 MeV$^{-1}$  \\
    $b$ & 0.0112371 MeV$^{-1}$ \\
    $c$ & 40 MeV \\
    $\omega$ & 5 MeV \\ [1ex]
    \hline
    \end{tabular}
    \caption{Parameters used for $^{16}$O to model the background $1s_{1/2}$ shell.}
    \label{table:background_shell_params}
\end{table}

\begin{table} [H]
    \centering
    \begin{tabular}{  c c c }
    \hline \hline
    Nucleus & p & n \\ [0.5 ex]
    \hline \hline
    $^{12}$C & 0.8 & 0.8\\
    $^{16}$O & 1.4 & 1.4\\
    $^{40}$Ar & 2.1 & 2.9\\ [1ex]
    \hline
    \end{tabular}
    \caption{Proton (p) and neutron (n) occupancies of the background $1s_{1/2}$ shell.}
    \label{table:background_shell_occup}
\end{table}

\section{}    
\label{sec:app:B}
\subsection{Hadronic tensor tables}

\noindent The tables are given as a seven-column text file with the hadronic tensor being a function of:
\begin{itemize}
	\item scattered nucleon kinetic energy, $T_{N}$;
	\item scattered nucleon polar angle, $\theta_{N}$;
	\item third component of the missing momentum, $p_{m}^{z}$;
	\item and missing energy, $E_{m}$.
\end{itemize}
The other three columns comprise the tensor index ($0$ to $10$, with the other $6$ being removed through symmetry arguments), real component and imaginary component of the tensor. The tables were calculated in the reference frame such that $\mathbf{q}$ is parallel to $\hat{\mathbf{z}}$. The kinematics are initially thrown in the laboratory frame in which the target nucleus is at rest. As a result, appropriate transformations are applied within the model to transform them into this hadronic tensor reference frame and back after the hadronic tensor value is retrieved. The tables are encoded such that only integers are stored for the above four bulleted variables. The integers describe steps in a given step size between a pre-defined minimum and maximum value. This was done to save memory when saving the tables to a text file. \\

\noindent The hadronic tensor is only precomputed for the aforementioned variables within a certain range. This range was chosen for two reasons:
\begin{enumerate}
	\item The T2K experiment flux tails off to a negligible value, while approximately 5\% of the DUNE flux is missed at the upper limit.
	\item Calculating the hadronic current takes time, and, for larger values of $T_{N}$, the computation time increases.
\end{enumerate}
The ranges of the aforementioned variables are given below.

\subsection{Ranges of variables used in tensor tables}

\noindent The ranges of variables used in the tables are given in Table~\ref{table:app:table_var_range}. The units for $T_{N}$ is MeV, $\theta_{N}$ is degrees and $p_{m}^{z}$ is MeV.

\begin{table} [H]
    \centering
    \begin{tabular}{  c c c c }
    \hline \hline
    Variable & Min & Max & Step \\ [0.5 ex]
    \hline \hline
    $T_{N}$ & 1 & 1101 & 10 \\
    $\theta_{N}$ & 0 & 180 & 2.0 \\
    $p_{m}^{z}$ & -800 & 800 & 20.0 \\ [1ex]
    \hline
    \end{tabular}
    \caption{Ranges of variables used in the hadronic tensor tables. Units are in MeV apart from nucleon angle which is in degrees.}
    \label{table:app:table_var_range}
\end{table}

\noindent It is important to note that the cross section changes quite rapidly around $T_N<10$ MeV. As a result, between $0$ and $10$ MeV, there is a separate table with a dedicated $1$ MeV $T_{N}$ step. All other variables are the same in this separate ``low $T_{N}$'' table. The $E_{m}$ variable is closely related to properties of individual nuclei and so the step size between nuclei changes.

\subsubsection{Carbon}
\noindent The $E_{m}$ range is shell dependent and neutrino helicity dependent. All following values in Table~\ref{table:app:c12_Em_prof} are given in MeV.

\begin{table} [H]
    \centering
    \begin{tabular}{ c c c c c }
    \hline \hline
    Neutrino helicity & Shell & Min & Max & Step \\ [0.5 ex]
    \hline \hline
    Neutrino & $1p_{3/2}$ & 17.8075 & 17.8075 & 1 \\
    Neutrino & $1s_{1/2}$ & 20 & 110 & 15 \\
    \hline
    Antineutrino & $1p_{3/2}$ & 14.46 & 14.46 & 1 \\
    Antineutrino & $1s_{1/2}$ & 20 & 110 & 15 \\
    \hline
    Both & Background & 20 & 300 & 40 \\
    \hline
    \end{tabular}
    \caption{Missing energy steps for $^{12}$C is different depending on the shell and neutrino helicity.}
    \label{table:app:c12_Em_prof}
\end{table}

\subsubsection{Oxygen}
\noindent For $^{16}$O, the $E_{m}$ range is also shell dependent and neutrino helicity dependent. All following values in Table~\ref{table:app:o16_Em_prof} are given in MeV.

\begin{table} [H]
    \centering
    \begin{tabular}{ c c c c c }
    \hline \hline
    Neutrino helicity & Shell & Min & Max & Step \\
    \hline \hline
    Neutrino & $1p_{1/2}$ & 15.4871 & 15.4871 & 1 \\
    Neutrino & $1p_{3/2}$ & 22.3652 & 22.3652 & 1 \\
    Neutrino & $1s_{1/2}$ & 20 & 110 & 15 \\
    \hline
    Antineutrino & $1p_{1/2}$ & 11.488 & 11.488 & 1 \\
    Antineutrino & $1p_{3/2}$ & 18.27 & 18.27 & 1 \\
    Antineutrino & $1s_{1/2}$ & 20 & 110 & 15 \\
    \hline 
    Both & Background & 20 & 300 & 40 \\
    \hline
    \end{tabular}
    \caption{Missing energy steps for $^{16}$O is different depending on the shell and neutrino helicity.}
    \label{table:app:o16_Em_prof}
\end{table}

\subsubsection{Argon}
\noindent For $^{40}$Ar, the same is true and all values given in Table~\ref{table:app:ar40_Em_prof} are in MeV.

\begin{table} [H]
    \centering
    \begin{tabular}{ c c c c c }
    \hline \hline
    Neutrino helicity & Shell & Min & Max & Step \\ 
    \hline \hline
    Neutrino & $2p_{1/2}$ & 9.857 & 9.857 & 1 \\
    Neutrino & $1d_{3/2}$ & 10.023 & 10.023 & 1 \\
    Neutrino & $1d_{5/2}$ & 10 & 40 & 10 \\
    Neutrino & $1p_{1/2}$ & 10 & 100 & 15 \\
    Neutrino & $1p_{3/2}$ & 10 & 100 & 15 \\
    Neutrino & $1s_{1/2}$ & 20 & 110 & 15 \\
    \hline 
    Antineutrino & $1f_{7/2}$ & 8.27 & 8.27 & 1 \\
    Antineutrino & $2s_{1/2}$ & 16.11 & 16.11 & 1 \\
    Antineutrino & $1d_{3/2}$ & 15.95 & 15.95 & 1 \\
    Antineutrino & $1d_{5/2}$ & 10 & 40 & 5 \\
    Antineutrino & $1p_{1/2}$ & 10 & 100 & 15 \\
    Antineutrino & $1p_{3/2}$ & 10 & 100 & 15 \\
    Antineutrino & $1s_{1/2}$ & 20 & 110 & 15 \\
    \hline 
    Both & Background & 20 & 300 & 40 \\
    \hline
    \end{tabular}
    \caption{Missing energy steps for $^{40}$Ar is different depending on the shell and neutrino helicity.}
    \label{table:app:ar40_Em_prof}
\end{table}

\subsection{Hadronic tensor table interpolation}

\noindent The hadronic tensor tables are interpolated using a four-dimensional bilinear interpolation. To estimate the value of a complex function \(f(x, y, z, w)\) at an arbitrary point \((p_x, p_y, p_z, p_w)\), the algorithm iteratively applies the linear interpolation formula across the four dimensions. For a single dimension, the linear interpolation between two points \(x_0\) and \(x_1\) is given by:  
\[
f_{\text{interp}}(p_x) = \frac{x_1 - p_x}{x_1 - x_0} f(x_0) + \frac{p_x - x_0}{x_1 - x_0} f(x_1).
\]  
Extending this to four dimensions, the interpolated value \(f_{\text{interpolated}}\) is calculated by first performing the interpolation over the \(x\)-dimension:
\[
f_x(y, z, w) = \frac{x_1 - p_x}{x_1 - x_0} f(x_0, y, z, w) + \frac{p_x - x_0}{x_1 - x_0} f(x_1, y, z, w).
\]  
This result is then interpolated along the \(y\)-dimension:
\[
f_{xy}(z, w) = \frac{y_1 - p_y}{y_1 - y_0} f_x(y_0, z, w) + \frac{p_y - y_0}{y_1 - y_0} f_x(y_1, z, w).
\]  
Next, interpolation proceeds in the \(z\)-dimension:
\[
f_{xyz}(w) = \frac{z_1 - p_z}{z_1 - z_0} f_{xy}(z_0, w) + \frac{p_z - z_0}{z_1 - z_0} f_{xy}(z_1, w).
\]  
Finally, interpolation along the \(w\)-dimension yields the fully interpolated value:
\[
f_{\text{interpolated}} = \frac{w_1 - p_w}{w_1 - w_0} f_{xyz}(w_0) + \frac{p_w - w_0}{w_1 - w_0} f_{xyz}(w_1).
\]  
In the implementation, the algorithm calculates the bounds \([x_0, x_1], [y_0, y_1], [z_0, z_1], [w_0, w_1]\) from the input values and extracts the corresponding function values from a four-dimensional array. This algorithm is applied to each hadronic tensor index.

\section{}    
\label{sec:app:C}
\subsection{Without nuclear recoil}
\noindent The Jacobian required is given as

\begin{equation}\label{app:eq:jacobian}
\Bigg | \frac{\partial p_{N}}{\partial E_{m}} \Bigg |.
\end{equation}

\noindent This can be expressed approximately by considering the definition of $E_{m}$ within the system. 

\begin{equation}\label{eq:pn2}
p_{N}^{2} = E_{N}^{2} - M_{N}^{2},
\end{equation}

\noindent where $E_{N}$ can be expressed as $T_{N} + M_{N}$. The missing energy is given as 

\begin{equation}
E_{m} = \omega - T_{N} - T_{B},
\end{equation}

\noindent where $T_{B}$ is the energy of the residual nucleus and $\omega$ is the energy transfer given by $E_{\nu} - E_{l}$. Using this, we can express $E_{m}$ as follows.

\begin{equation}
E_{m} = \omega - E_{N} + M_{N} - T_{B}.
\end{equation} 

\noindent This can be substituted into Eq.~\ref{eq:pn2} to yield

\begin{equation}
p_{N}^{2} = (\omega - E_{m} + M_{N} - T_{B})^{2} - M_{N}^{2}.
\end{equation}

\noindent Performing the required differentiation in Eq.~\ref{eq:jacobian}, 

\begin{equation}
2p_{N} \frac{\partial p_{N}}{\partial E_{m}} = -2(\omega - E_{m} + M_{N} - T_{B}) = -2E_{N}.
\end{equation}

\noindent Finally, the required Jacobian is 

\begin{equation}
\Bigg | \frac{\partial p_{N}}{\partial E_{m}} \Bigg | = \frac{E_{N}}{p_{N}}.
\end{equation}

\subsection{With nuclear recoil}

\noindent To obtain the Jacobian given in~\ref{app:eq:jacobian}, we make use of the relation

\begin{equation}\label{app:eq:jacobian-expansion}
 \frac{\partial E_{m}}{\partial p_{N}} = \frac{\partial E_{m}}{\partial E_{B}} \frac{\text{d} E_{B}}{\text{d} p_{N}} + \frac{\partial E_{m}}{\partial E_{N}} \frac{\text{d} E_{N}}{\text{d} p_{N}}.
\end{equation}

\noindent where $E_{B}$ is the recoil energy. The recoil energy and other useful relations are given below in Eq.~\ref{app:eq:relations}

\begin{equation}\label{app:eq:relations}
    \begin{split}
        & E_{B}^{2} = M_{B}^{2} + p_{B}^{2} ,\\
        & p_{B}^{2} = q^{2} + p_{N}^{2} - 2qp_{N}\cos(\theta_{qN}) ,\\
        & M_{B} = E_{m} - E_{N} + M_{A}, \\
        & E_{N} = M_{A} + \omega -  \\ &\sqrt{q^{2} + p_{N}^{2} - 2qp_{N}\cos(\theta_{qN}) + (E_{m} -M_{N} + M_{A})^{2}}.
    \end{split}
\end{equation}

\noindent Here, the subscript ``$B$'' refers to the residual nucleus, $q$ is the magnitude of the momentum transfer and $\theta_{qN}$ is the angle between the momentum transfer and the scattered nucleon. The first term in Eq.~\ref{app:eq:jacobian-expansion} is 

\begin{equation}
\frac{\partial E_{m}}{\partial E_{B}}\frac{\text{d} E_{B}}{\text{d} p_{N}}  = \frac{E_{B}}{M_{B}} \Bigg(\frac{p_{N} - q\cos(\theta_{qN})}{E_{B}} \Bigg),
\end{equation}

\noindent and the second term is 
\begin{equation}
\frac{\partial E_{m}}{\partial E_{N}}\frac{\text{d} E_{N}}{\text{d} p_{N}}  = \frac{E_{B}}{M_{B}} \Bigg(\frac{p_{N}}{E_{N}} \Bigg).
\end{equation}

\noindent The final Jacobian is then given as the inverse of the above

\begin{equation}
\begin{split}   
\Bigg | \frac{\partial p_{N}}{\partial E_{m}} \Bigg | &=\Bigg | \frac{\partial E_{m}}{\partial p_{N}} \Bigg |^{-1}\\
& = \frac{M_{B}}{E_{B}} \Bigg| \frac{p_{N}}{E_{N}} + \frac{p_{N} - q\cos(\theta_{qN})}{E_{B}} \Bigg|^{-1}
\end{split}
\end{equation}

\section{}    
\label{sec:app:D}

\subsection{Total cross section and maximum differential cross section}
\begin{figure}[H]
    \centering
    \includegraphics[width=0.5\textwidth]{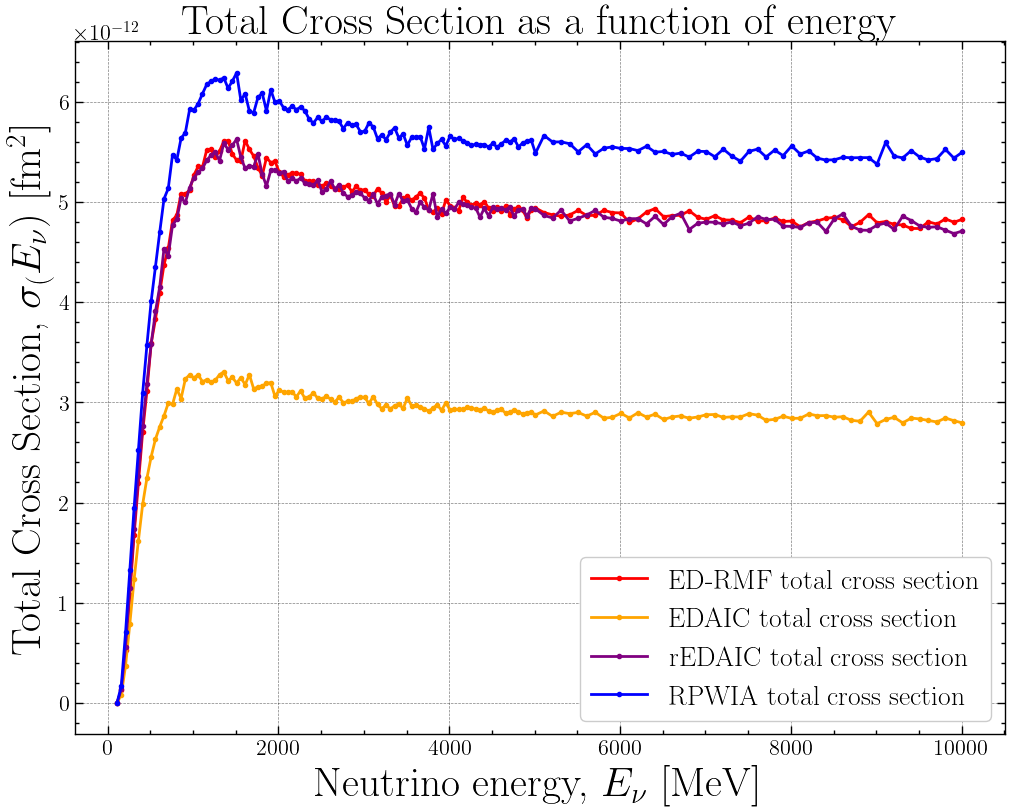}
    \caption{Total cross section as a function of neutrino energy for $\nu_{\mu}$ on $^{12}$C. The models shown are EDRMF, ROP (EDAIC), real only ROP (rEDAIC) and relativistic plane wave impulse approximation (RPWIA). }
    \label{fig:total-xsec-all-models}
\end{figure}

\noindent The NEUT framework, when used with a neutrino flux, requires a total cross section as a function of neutrino energy. NEUT obtains an event rate from the composition of the flux with the total cross section. The total cross section table is required for all hadronic tables. Fig.~\ref{fig:total-xsec-all-models} shows the total cross section as a function of neutrino energy for different nuclear potentials.\\

\noindent The total cross section is calculated by systematically stepping through incident neutrino energy and calculating the six-fold differential cross section for all events and then summing the six-fold differential cross section, which is then divided by the total number of events thrown for that given neutrino energy. This value is then multiplied by the total averaged six-dimensional phase space to obtain a total cross section for that neutrino energy. The maximum of the six-fold differential cross section values is retained for each neutrino energy and used as a ceiling with which to throw a random value of the six-fold differential cross section. A new maximum differential cross section is required for each neutrino helicity and nuclear potential model. \\

\section{}    
\label{sec:app:E}
\subsection{Comparison with pre-correction MINER$\nu$A data}
\begin{figure*}
    \begin{minipage}[h!]{\textwidth}
    \includegraphics[height=15cm, width=0.75\textwidth]{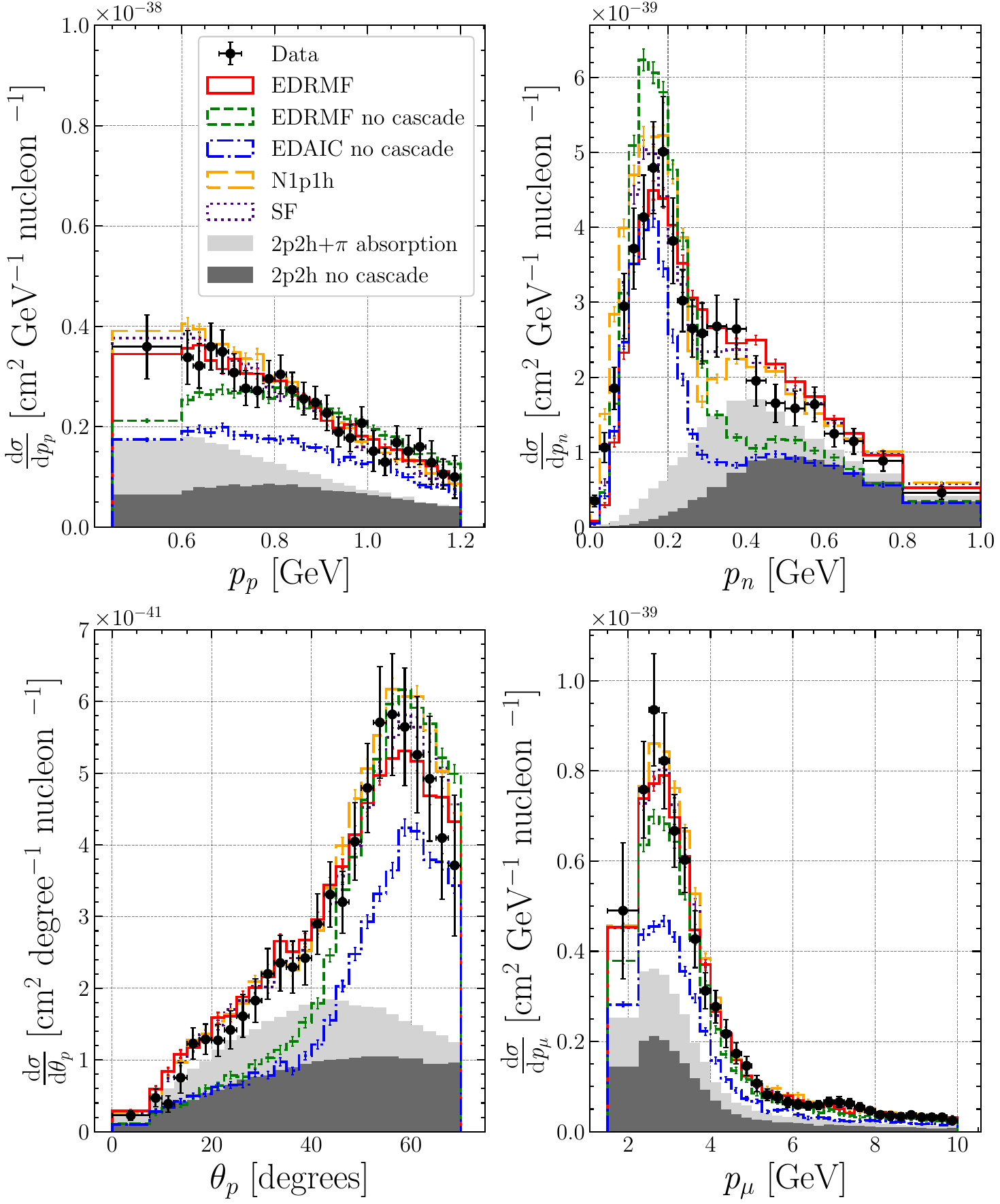}
    \caption{MINER$\nu$A $\nu_{\mu}$ CC$0\pi$N$p$ dataset on hydrocarbon. Differential cross sections for different kinematics are shown. Subscript $p$ indicated a scattered proton kinematic; a subscript $\mu$ indicates a scattered lepton kinematic and a subscript $n$ indicates a reconstructed initial neutron kinematic. The histograms follow the same definition as Fig.~\ref{fig:t2k2018:1p_costheta}. Data taken from~\cite{MinervaData:2018}.}
    \label{fig:minerva:kinematics-old}
    \end{minipage}
\end{figure*}

\begin{figure*}
    \begin{minipage}[h!]{\textwidth}
    \includegraphics[height=15cm, width=0.75\textwidth]{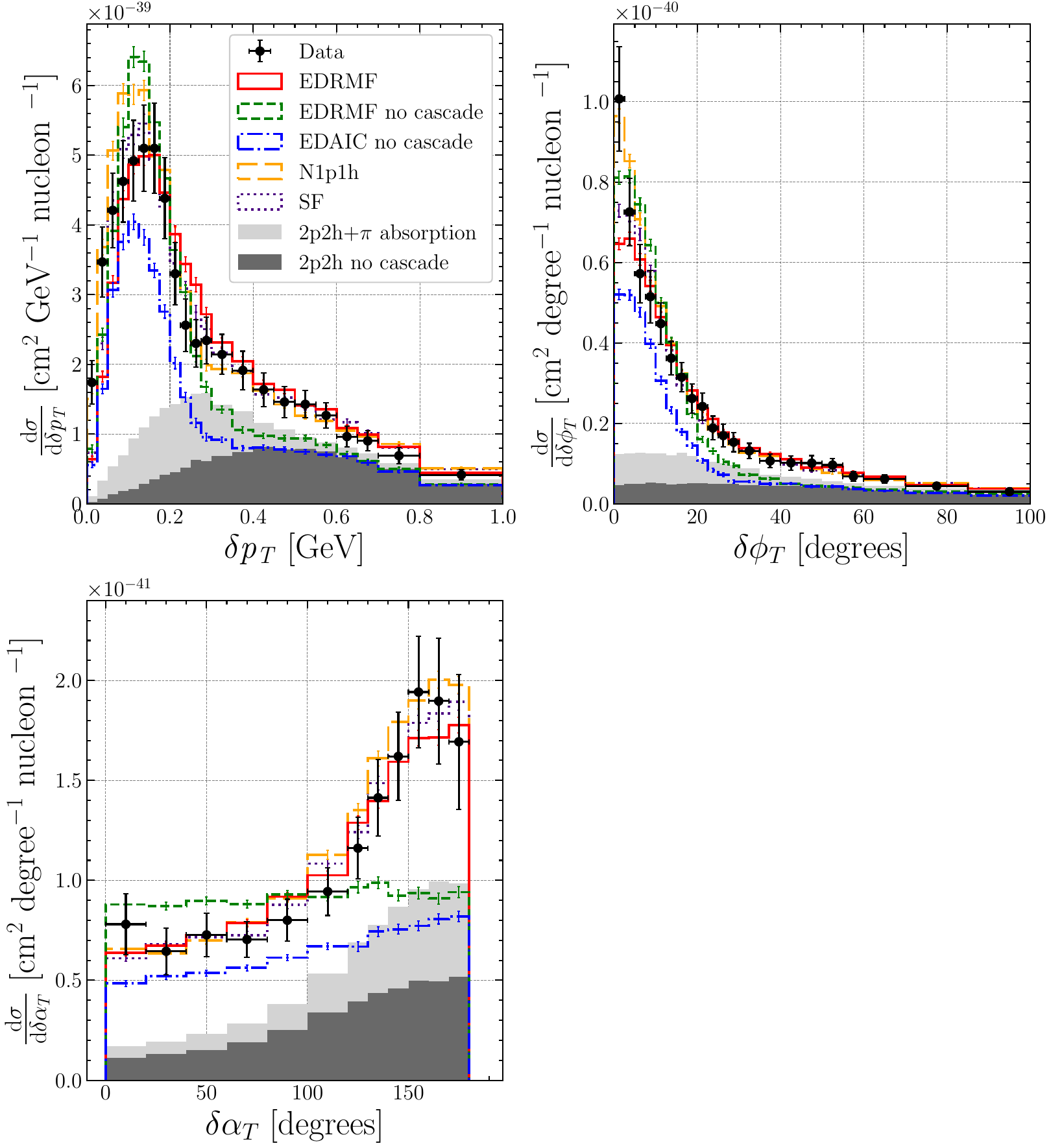}
    \caption{MINER$\nu$A $\nu_{\mu}$ CC$0\pi$N$p$ dataset on hydrocarbon. Differential cross sections for TKI variables defined in Eq.~\ref{eq:tki}. The histograms follow the same definition as Fig.~\ref{fig:t2k2018:1p_costheta}. Data taken from~\cite{MinervaData:2018}.}
    \label{fig:minerva:TKI-old}
    \end{minipage}
\end{figure*}

\noindent
The correction applied in Ref.~\cite{minerva-data-set} particularly influences low bins in $\delta p_{T}$, $\delta \phi_{T}$ and $p_{n}$. The comparison to the dataset before the correction, given in Fig.~\ref{fig:minerva:kinematics-old} and \ref{fig:minerva:TKI-old} with the respective $\chi^{2}$ values shown in Tables~\ref{table:minerva_kinematic_chi2-old} and \ref{table:minerva_tki_chi2-old}, is shown to highlight the effect of comparing to the updated dataset.

\begin{table} [H]
    \centering
    \begin{tabular}{  c  | c   c   c  c }
    \hline \hline
    Model & $p_{p}$ & $p_{n}$ & $\theta_{p}$ & $p_{\mu}$  \\ [0.5 ex]
    \hline \hline
    EDRMF cas & $29.2/25$ & $63.1/24$ & $47.8/26$ & $49.1/32$\\
    EDRMF no cas & $40.2/25$ & $164/24$ & $61.3/26$ & $44.8/32$\\
    N1p1h cas & $42.6/25$ & $131/24$ & $61.4/26$ & $40.8/32$\\
    SF cas & $43.5/25$ & $73.7/24$ & $36.5/26$ & $48.6/32$\\
    EDAIC no cas & $33.9/25$ & $65.9/24$ & $44.0/26$ & $39.5/32$\\ [1ex]
    \hline
    \end{tabular}
    \caption{$\chi^{2}/N_{dof}$ values for each kinematic variable variable for MINER$\nu$A. ``cas'' and ``no cas'' indicate where the cascade has and has not been applied respectively.
    }
    \label{table:minerva_kinematic_chi2-old}
\end{table}

\begin{table} [H]
    \centering
    \begin{tabular}{  c |  c   c   c  }
    \hline \hline
    Model & $\delta p_{T}$ & $\delta \phi_{T}$ & $\delta \alpha_{T}$  \\ [0.5 ex]
    \hline \hline
    EDRMF cas & $61.8/24$ & $73.8/23$ & $21.7/12$\\
    EDRMF no cas & $167/24$ & $112/23$ & $53.1/12$\\
    N1p1h cas & $177/24$ & $89.5/23$ & $17.0/12$ \\
    SF cas & $201/24$ & $93.3/23$ & $19.5/12$\\
    EDAIC no cas & $98.0/24$ & $64.6/23$ & $38.0/12$ \\ [1ex]
    \hline
    \end{tabular}
    \caption{$\chi^{2}/N_{dof}$ values for each TKI variable for MINER$\nu$A. ``cas'' and ``no cas'' indicate where the cascade has and has not been applied respectively.
    }
    \label{table:minerva_tki_chi2-old}
\end{table}

\clearpage

\section{}    
\label{sec:app:F}
\noindent Shape-only comparisons are produced by NUISANCE~\cite{Nuisance}. 
Scale factors required to scale the overall normalisation of the MC are calculated by taking the ratio of the data distribution with that of the MC distribution. A scale factor close to 1.0 with a good shape agreement by eye indicates a good agreement with the data. A good shape with a scale factor that is not 1.0 indicates good agreement in shape but a difference in normalisation.
\subsection{Shape-only comparison for T2K}
\begin{figure*}
    \begin{minipage}[h!]{\textwidth}
    \includegraphics[keepaspectratio, width=0.85\textwidth]{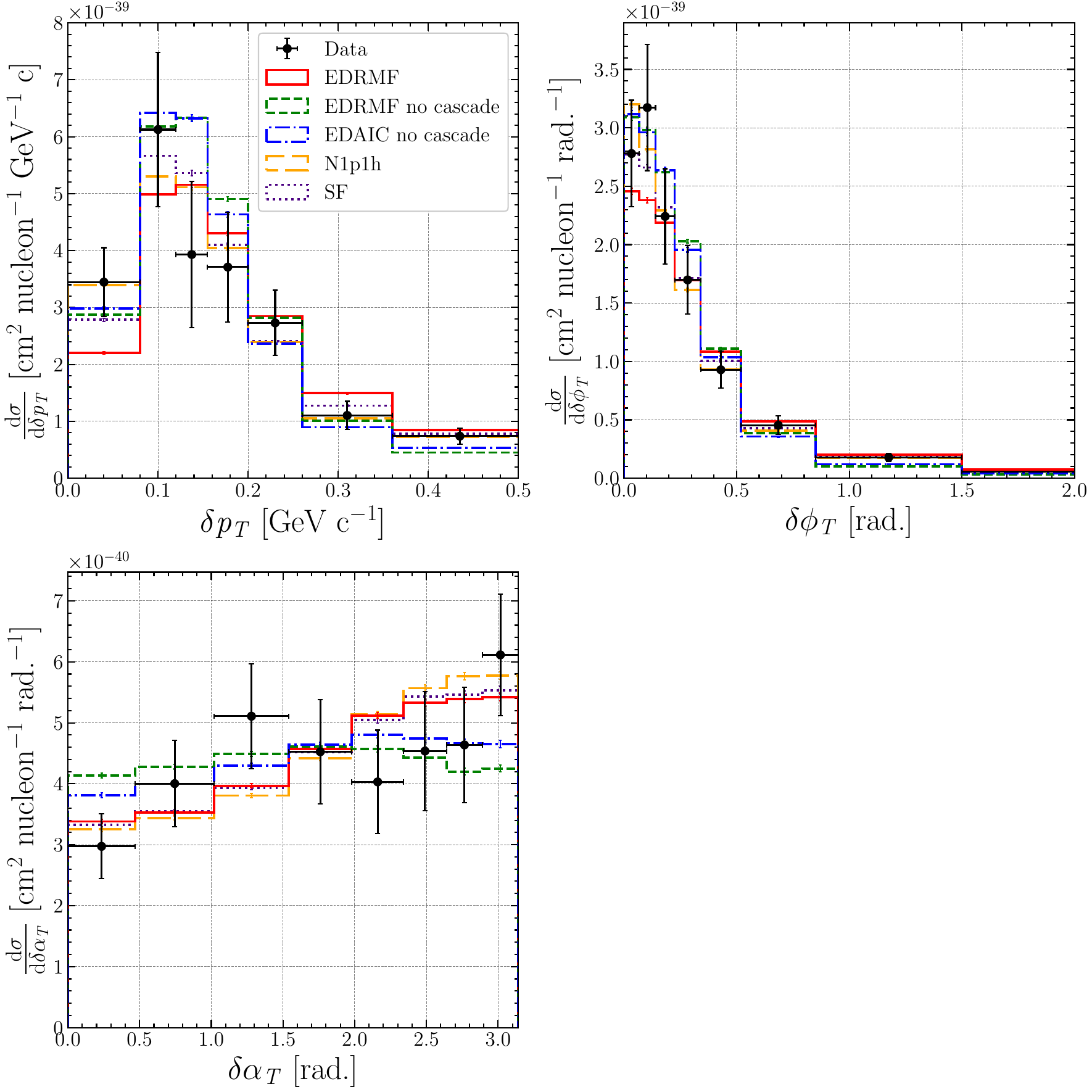}
    \caption{Shape-only differential cross sections for TKI variables defined in Eq.~\ref{eq:tki} for T2K. The NEUT models are normalised to the measurement. The histograms follow the same definition as Fig.~\ref{fig:t2k2018:1p_costheta}. Data taken from~\cite{T2K_2018_data}.}
    \label{fig:T2K:TKI:shape}
    \end{minipage}
\end{figure*}

\noindent Fig.~\ref{fig:T2K:TKI:shape} shows a TKI shape-only comparison of the considered NEUT models and the measurement. The corresponding scale factors required to scale the MC to data are given in Table~\ref{table:T2K:TKI:shape}. For $\delta \phi_{T}$, all models show agreement from approximately $0.4$ rad. At lower values of $\delta \phi_{T}$, the EDRMF model with the cascade is the lowest in strength. For $\delta \alpha_{T}$, models with the cascade applied display a similar shape; likewise for models without the cascade. The EDRMF model without the cascade and the EDAIC model have similar shapes for all three TKI variables.

\begin{table}[h]
\centering
\begin{tabular}{c|c c c }
    \hline \hline
Model & $\delta p_{T}$ & $\delta \phi_{T}$ & $\delta \alpha_{T}$ \\
\hline \hline
EDRMF cas & 1.00 & 1.02 & 1.05 \\
EDRMF no cas & 0.92 & 0.94 & 0.97 \\
N1p1h cas & 0.88 & 0.89 & 0.92 \\
SF cas & 0.95 & 0.97 & 1.00 \\
EDAIC no cas & 1.37 & 1.39 & 1.44 \\

\hline
\end{tabular}
\caption{    
Scale factors requred to scale the MC to data for each TKI variable for T2K. ``cas'' and ``no cas'' indicate where the cascade has and has not been applied respectively.}
\label{table:T2K:TKI:shape}
\end{table}

\subsection{Shape-only comparison for MINER$\nu$A}
\begin{figure*}
    \begin{minipage}[h!]{\textwidth}
    \includegraphics[keepaspectratio, width=0.85\textwidth]{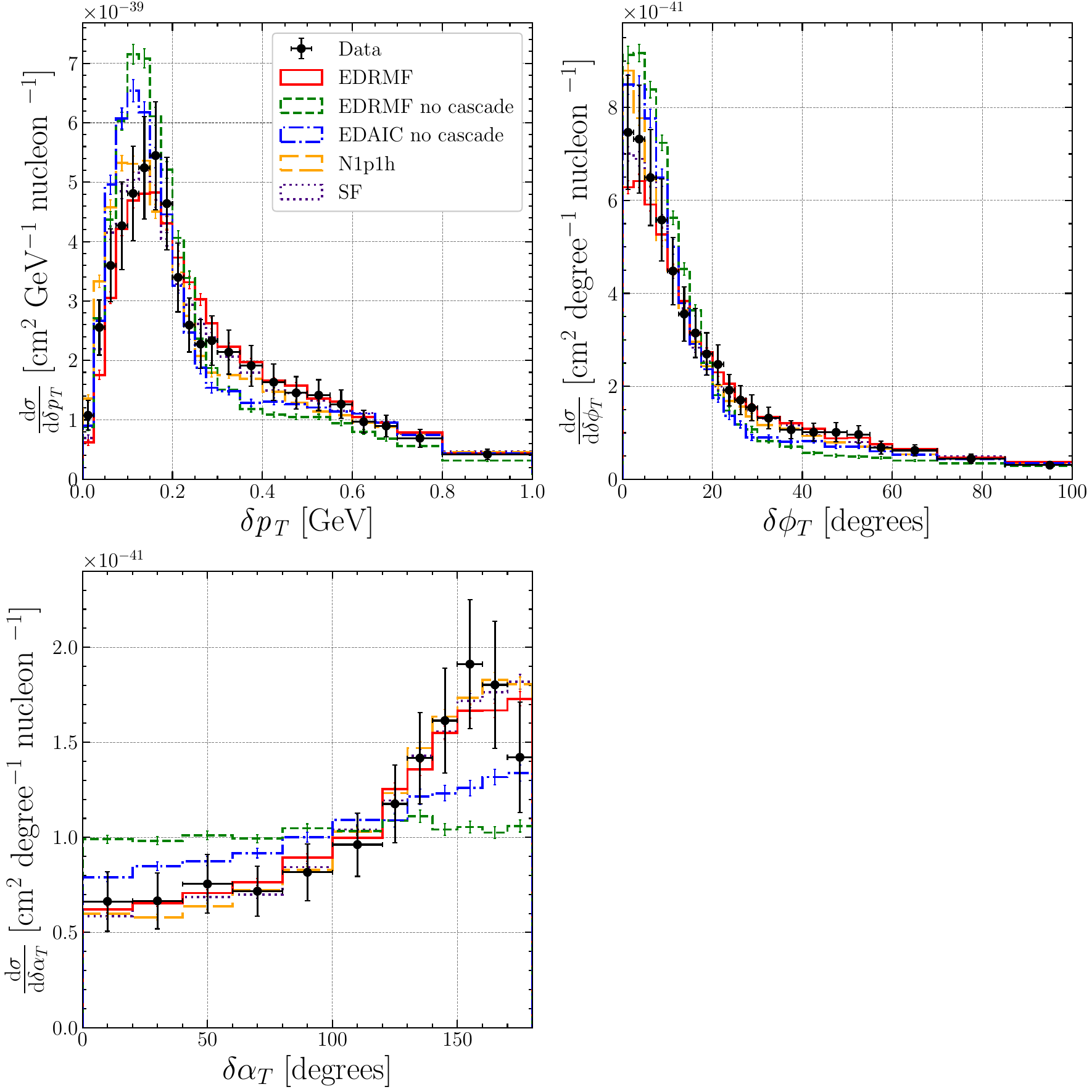}
    \caption{Shape-only differential cross sections for TKI variables defined in Eq.~\ref{eq:tki} for MINER$\nu$A. The NEUT models are normalised to the measurement. The histograms follow the same definition as Fig.~\ref{fig:t2k2018:1p_costheta}. Data taken from~\cite{minerva-data-set}.}
    \label{fig:MINERvA:TKI:shape}
    \end{minipage}
\end{figure*}

\noindent Fig.~\ref{fig:MINERvA:TKI:shape} shows a TKI shape-only comparison of the considered NEUT models and the measurement. The corresponding scale factors required to scale the MC to data are given in Table~\ref{table:MINERvA:TKI:shape}. For $\delta p_{T}$, the best shape agreement by eye is the SF model. For $\delta \phi_{T}$, all models with the cascade agree beyond around 10 degrees. The value of $\delta \phi_{T}$ at 0 degrees differs dramatically between the three models that employ the cascade. For $\delta \alpha_{T}$, all models with the cascade applied agree until around 100 degrees, differing at higher angles. In the case of $\delta \alpha_{T}$, the EDRMF model and the EDAIC model without the cascade, do not agree in shape.  \\

\begin{figure*}
    \begin{minipage}[h!]{\textwidth}
    \includegraphics[keepaspectratio, width=0.85\textwidth]{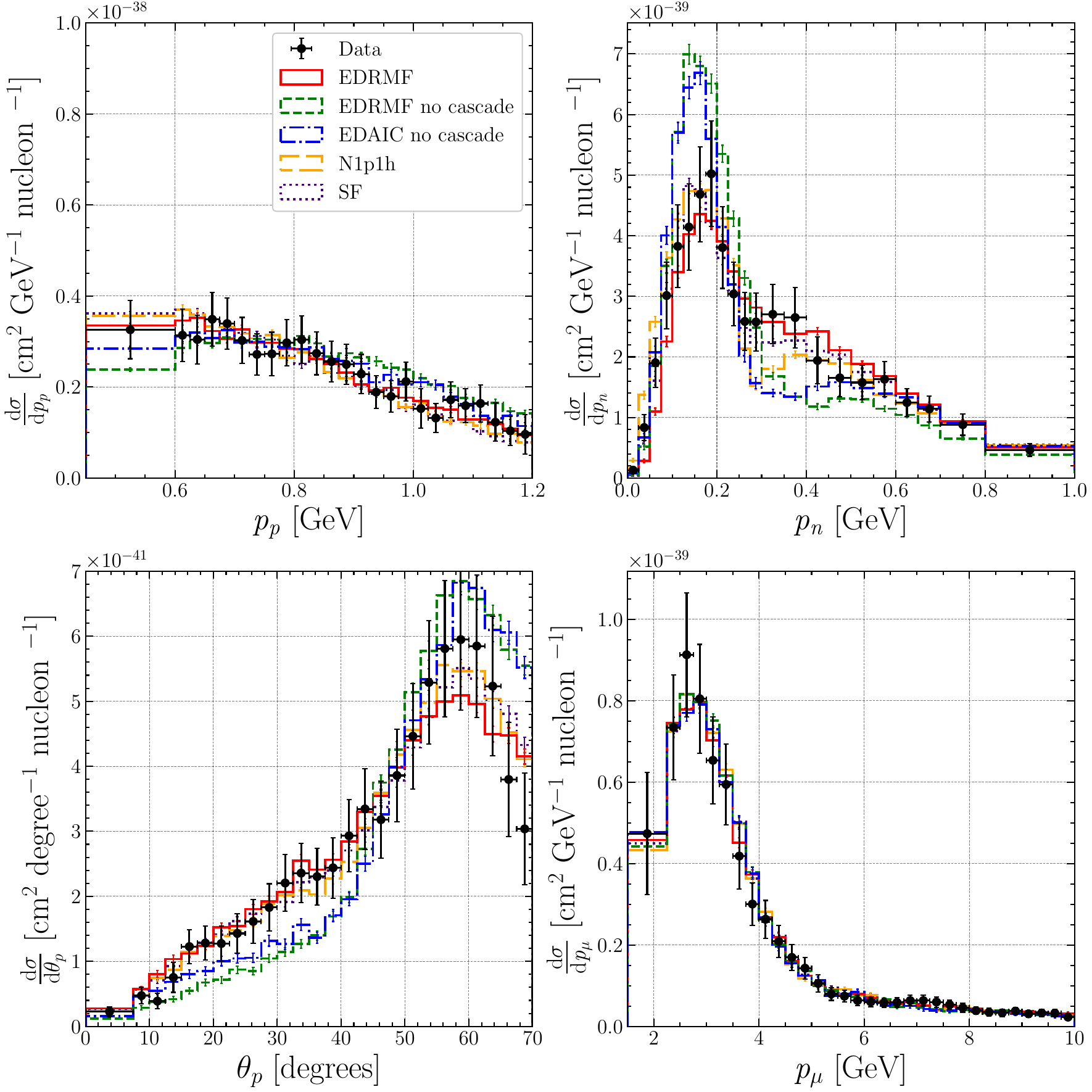}
    \caption{Shape-only differential cross sections for kinematic variables for MINER$\nu$A. The NEUT models are normalised to the measurement. The histograms follow the same definition as Fig.~\ref{fig:t2k2018:1p_costheta}. Data taken from~\cite{minerva-data-set}.}
    \label{fig:MINERvA:kinematics:shape}
    \end{minipage}
\end{figure*}

\noindent Fig.~\ref{fig:MINERvA:kinematics:shape} shows a shape-only comparison for the kinematic cross sections for the considered NEUT models and the measurement. The corresponding scale factors required to scale the MC to data are given in Table~\ref{table:MINERvA:kinematics:shape}. For $p_{n}$, the peak is captured well by the SF and N1p1h models; however, the second peak around 0.3-0.4 GeV is captured only by the EDRMF model with the cascade applied. For all variables, the EDRMF model without the cascade and the EDAIC model agree in shape. \\

\noindent For $p_{\mu}$, all models closely match the shape of the data. The peak is not captured by any of the models but is consistent within the error bar. This indicates that, in terms of cross section shape, all models differ mainly in the description of the hadronic part of the interaction. For $\theta_{p}$, all models with the cascade capture the shape at low angles, but at high angles the strength is lower, although consistent within the error.

\begin{table}[h]
\centering
\begin{tabular}{c|c c c }
    \hline \hline
Model & $\delta p_{T}$ & $\delta \phi_{T}$ & $\delta \alpha_{T}$ \\
\hline \hline
EDRMF cas & 0.96 & 0.97 & 0.97 \\
EDRMF no cas & 1.12 & 1.13 & 1.13 \\
N1p1h cas & 0.90 & 0.91 & 0.91 \\
SF cas & 0.95 & 0.96 & 0.96 \\
EDAIC no cas & 1.62 & 1.63 & 1.63 \\
\hline
\end{tabular}
\caption{    
Scale factors required to scale the MC to data for each TKI variable for MINER$\nu$A. ``cas'' and ``no cas'' indicate where the cascade has and has not been applied respectively.}
\label{table:MINERvA:TKI:shape}
\end{table}

\begin{table}[h]
\centering
\begin{tabular}{c|c c c c }
    \hline \hline
Model & $p_{p}$ & $p_{n}$ & $\theta_{p}$ & $p_{\mu}$ \\
\hline \hline
EDRMF cas & 0.97 & 0.97 & 0.96 & 1.01 \\
EDRMF no cas & 1.13 & 1.12 & 1.11 & 1.17 \\
N1p1h cas & 0.91 & 0.91 & 0.90 & 0.95 \\
SF cas & 0.96 & 0.96 & 0.95 & 1.00 \\
EDAIC no cas & 1.63 & 1.62 & 1.61 & 1.69 \\
\hline
\end{tabular}
\caption{    
Scale factors required to scale the MC to data for each kinematic variable variable for MINER$\nu$A. ``cas'' and ``no cas'' indicate where the cascade has and has not been applied respectively.
}
\label{table:MINERvA:kinematics:shape}
\end{table}

\subsection{Shape-only comparison for MicroBooNE}

\begin{figure*}
    \begin{minipage}[h!]{\textwidth}
    \includegraphics[keepaspectratio, width=0.85\textwidth]{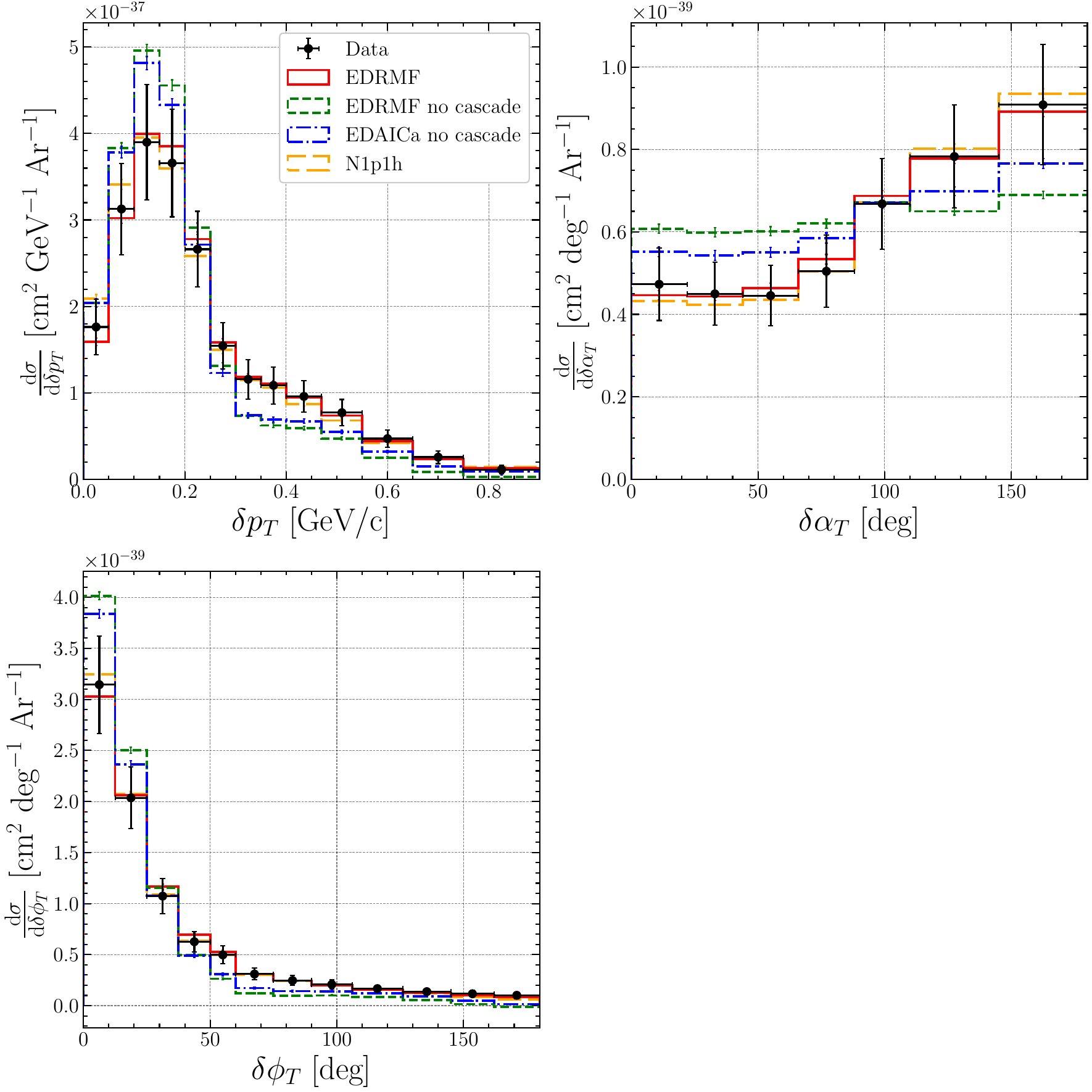}
    \caption{Shape-only differential cross sections for TKI variables defined in Eq.~\ref{eq:tki} for MicroBooNE. The NEUT models are normalised to the measurement. The histograms follow the same definition as Fig.~\ref{fig:t2k2018:1p_costheta} except the SF model is not included. Data taken from~\cite{uboone-TKI:2023, uboone-kinematics:2020}.}
    \label{fig:uboone:TKI:shape}
    \end{minipage}
\end{figure*}

\noindent Fig.~\ref{fig:uboone:TKI:shape} shows a shape-only comparison of the considered NEUT models and the measurement. The corresponding scale factors required to scale the MC to data are given in Table~\ref{table:uboone_tki_scalef:shape}. For all TKI variables, the N1p1h and EDRMF model with the cascade agree to the data well and have almost identical shape. For $\delta p_{T}$ and $\delta \phi_{T}$ it is interesting to note the agreement in shape for the EDRMF model without the cascade and the EDAICa model, indicating that they differ mostly in normalisation. \\

\begin{figure*}
    \begin{minipage}[h!]{\textwidth}
    \includegraphics[keepaspectratio, width=0.85\textwidth]{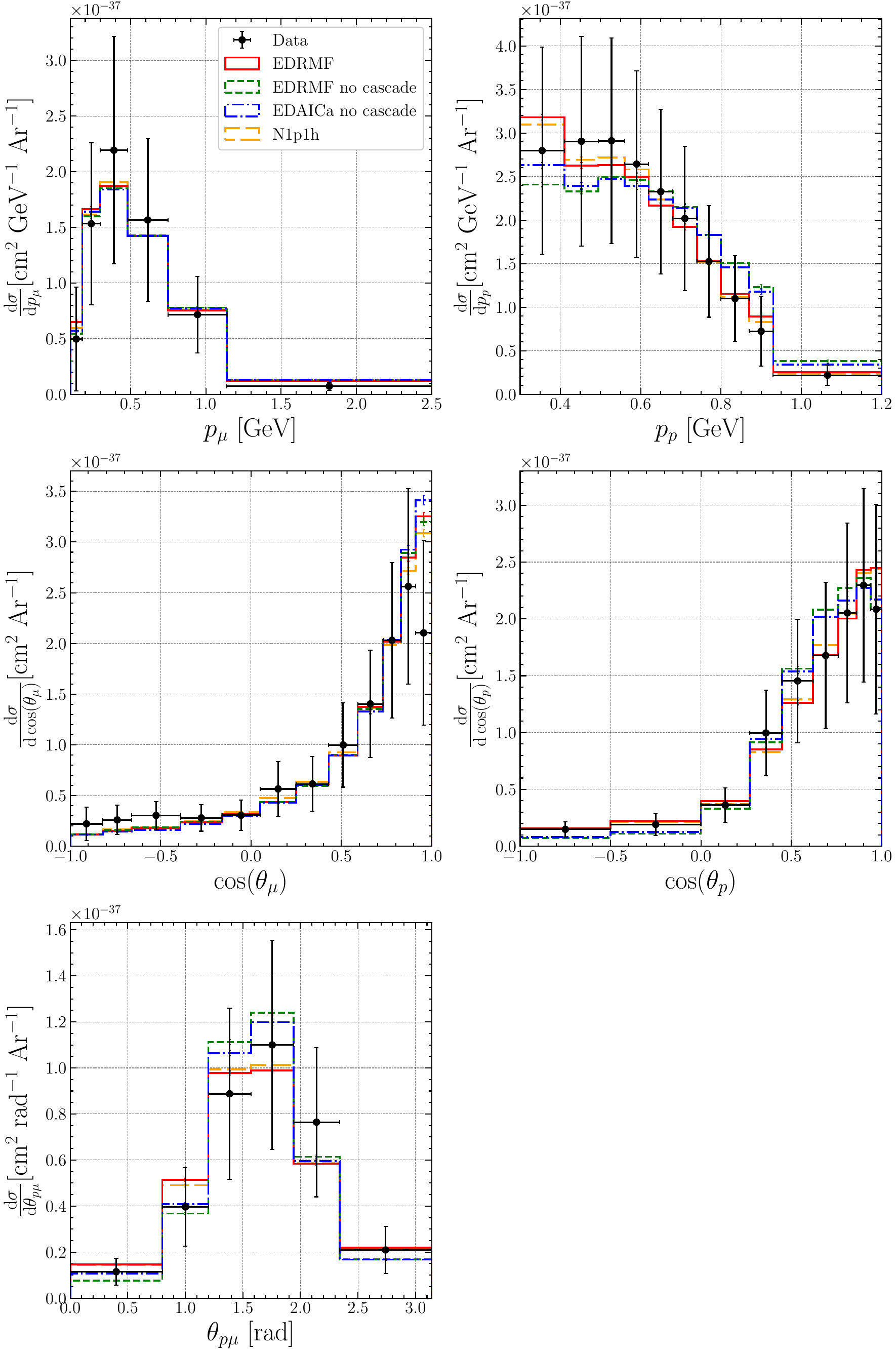}
    \caption{Shape-only differential cross sections for kinematic variables for MicroBooNE. The NEUT models are normalised to the measurement. The histograms follow the same definition as Fig.~\ref{fig:t2k2018:1p_costheta} except the SF model is not included. Data taken from~\cite{uboone-kinematics:2020}.}
    \label{fig:uboone:kinematics:shape}
    \end{minipage}
\end{figure*}

\noindent Fig.~\ref{fig:uboone:kinematics:shape} shows a shape-only comparison for the kinematic cross sections for the considered NEUT models and the measurement. The corresponding scale factors required to scale the MC to data are given in Table~\ref{table:uboone:kinematics:shape}. For $p_{\mu}$, all models agree in shape and are consistent with the data. A similar agreement is seen in $\cos(\theta_{\mu})$, however, small shape disagreements are seen at $\cos(\theta_{\mu}) \approx 1.0$, indicating that at forward muon scattering angles, each model describes a different angular distribution. All models overestimate the final bin, where the measurement shows a drop. \\

\noindent For $p_{p}$, the EDRMF model with the cascade and the N1p1h model agrees in shape and is consistent with data. The EDRMF model without the cascade and the EDAICa model have similar shapes but have lower strength at lower proton momentum and higher strength at higher proton momentum. Similarly, for $\cos(\theta_{p})$, the EDRMF model with the cascade and the N1p1h model agree in shape and with the data. The models without the cascade applied have slightly lower strength at lower $\cos(\theta_{p})$ and slightly higher strength between $0.25 < \cos(\theta_{p}) < 0.8$ before having lower strength again in the final bin. \\

\noindent For $\theta_{p\mu}$, between $0$ and $1$ radians, models without the cascade have a better shape by eye before having a greater strength between $1$ and $2$ radians. The EDRMF model with the cascade and the N1p1h model agree in shape and are consistent with the data.

\begin{table}[h]
\centering
\begin{tabular}{c|c c c }
    \hline \hline
Model & $\delta p_{T}$ & $\delta \phi_{T}$ & $\delta \alpha_{T}$ \\
\hline \hline
EDRMF cas & 1.40 & 1.36 & 1.39 \\
EDRMF no cas & 1.12 & 1.21 & 1.12 \\
N1p1h cas & 1.31 & 1.40 & 1.36 \\
EDAICa no cas & 2.07 & 2.23 & 2.09 \\
\hline
\end{tabular}
\caption{    
Scale factors required to scale MC to data for each TKI variable for MicroBooNE. ``cas'' and ``no cas'' indicate where the cascade has and has not been applied respectively.}
\label{table:uboone_tki_scalef:shape}
\end{table}

\begin{table} [H]
    \centering
    \begin{tabular}{  c |  c   c   c  c  c }
    \hline \hline
    Model & $p_{\mu}$ & $p_{p}$ & $\cos(\theta_{\mu})$ & $\cos(\theta_{p})$ & $\theta_{p \mu}$ \\ [0.5 ex]
    \hline \hline
    EDRMF cas & 0.92 & 0.92 & 0.94 & 0.93 & 0.94 \\
    EDRMF no cas & 1.01 & 1.01 & 1.04 & 1.02 & 1.03 \\
    N1p1h cas & 0.88 & 0.88 & 0.90 & 0.89 & 0.90 \\
    EDAICa no cas & 1.79 & 1.79 & 1.83 & 1.80 & 1.82 \\[1ex]
    \hline
    \end{tabular}
    \caption{Scale factors required to scale the MC to data for each kinematic variable variable for MINER$\nu$A. ``cas'' and ``no cas'' indicate where the cascade has and has not been applied respectively. 
    }
    \label{table:uboone:kinematics:shape}
\end{table}

\end{appendices}

\end{document}